\documentclass[aps,twocolumn,showpacs,preprintnumbers,amsmath,amssymb,superscriptaddress,prl,nofootinbib]{revtex4-1}

\usepackage{graphicx}
\usepackage{dcolumn}
\usepackage{bm}
\usepackage{graphics}
\usepackage{subfigure}
\usepackage{graphicx}
\usepackage{epsfig}
\usepackage{epstopdf}
\usepackage{color}

\begin{document}

\title{Surface plasmon polaritons in topological Weyl semimetals}

\author{Johannes Hofmann}
\affiliation{Condensed Matter Theory Center and Joint Quantum Institute, Department of Physics, University of Maryland, College Park, Maryland 20742-4111 USA}
\affiliation{T.C.M. Group, Cavendish Laboratory, University of Cambridge, Cambridge CB3 0HE, United Kingdom}

\author{Sankar Das Sarma}
\affiliation{Condensed Matter Theory Center and Joint Quantum Institute, Department of Physics, University of Maryland, College Park, Maryland 20742-4111 USA}

\date{\today}

\begin{abstract}
We consider theoretically surface plasmon polaritons in Weyl semimetals. These materials contain pairs of band touching points -- Weyl nodes -- with a chiral topological charge, which induces an optical anisotropy and anomalous transport through the chiral anomaly. We show that these effects, which are not present in ordinary metals, have a direct fundamental manifestation in the surface plasmon dispersion. The retarded Weyl surface plasmon dispersion depends on the separation of the Weyl nodes in energy and momentum space. For Weyl semimetals with broken time-reversal symmetry, the distance between the nodes acts as an effective applied magnetic field in momentum space, and the Weyl surface plasmon polariton dispersion is strikingly similar to magnetoplasmons in ordinary metals. In particular, this implies the existence of nonreciprocal surface modes. In addition, we obtain the nonretarded Weyl magnetoplasmon modes, which acquire an additional longitudinal magnetic-field dependence. These predicted surface plasmon results are observable manifestations of the chiral anomaly in Weyl semimetals and might have technological applications.
\end{abstract}


%73.20.Mf	Electron states at surfaces and interfaces, Collective excitations (including excitons, polarons, plasmons and other charge-density excitations)
%78.68.+m	Optical properties of surfaces
%71.20.Gj		Electronic structure of semimetals
%03.65.Vf		Phases: geometric; dynamic or topological
\maketitle

Surface plasmon polaritons (SPPs) are collective electromagnetic and electron-charge excitations that are confined to the surface of a metal or semiconductor. They were proposed in the 1950's~\cite{ritchie57,stern60} and have been observed via electron energy loss spectroscopy~\cite{powell59,garcia10} as well as optically via surface gratings~\cite{ritchie68} or attenuated total reflection~\cite{marschall71}. Over the past decades, SPPs have found widespread technological applications, for example, in surface microscopy~\cite{rothenhausler88}, for biomolecular detection~\cite{malmqvist93}, or lithography~\cite{srituravanich04}. Because SPPs are focused to sizes smaller than the wavelength of light, they hold promise to realize miniaturized plasmon-based optoelectronic devices, and research in creating such plasmonic devices is flourishing~\cite{barnes03}, with the subject being dubbed ``plasmonics'' or ``nano plasmonics,'' which is a huge applied physics field in its own right.

In this Rapid Communication, we add a fundamental physical aspect to the study of SPPs (and the field of plasmonics), and demonstrate that the surface plasmon polaritons of recently discovered Weyl semimetals (WSMs), which possess topological properties, show a much richer (and unanticipated) structure compared to standard SPPs in ordinary metals and semiconductors. We find that due to the quantum anomalous electrodynamic response of the WSM (which is their hallmark), the retarded Weyl surface plasmon is strongly sensitive to details of the band structure. In particular, we find a geometry in which the SPP is nonreciprocal (i.e., the propagation is unidirectional), even without an applied external magnetic field. In addition, we show that the magnetoplasmon mode displays an additional longitudinal field dependence which is absent in ordinary metals. This can serve as a direct signature of Weyl semimetals in surface measurements. We note that the SPP physics introduced in this work applies to extrinsic or doped WSM materials with no requirement of fine-tuning the chemical potential to the band touching points, making our predictions easy to test experimentally.

An important aspect of SPPs for technological applications is their nonreciprocity, i.e., the SPPs can only propagate in one direction~\cite{barnes03}. In conventional metals, nonreciprocal modes are only possible by breaking time-reversal symmetry in an applied external magnetic field~\cite{chiu72,wallis74}. This comes with great  technological challenges since for a sizable nonreciprocity, these magnetic fields have to be very large~\cite{barnes03}. In this Rapid Communication, we report nonreciprocal SPPs in the pristine WSMs that are induced by topological Weyl node separation without any external magnetic field. This provides an alternative route to nonreciprocal modes and could point to interesting technological applications of WSMs in nanoplasmonics.

Weyl semimetals contain a valence and conduction band that touch in isolated points of the Brillouin zone near the chemical potential $\mu$. The minimal Hamiltonian in the vicinity of such a Weyl node is~\cite{vafek14,wehling14}
\begin{equation}
 H = \chi v {\bf p} \cdot \boldsymbol{\sigma} - \mu , \label{eq:HamiltonianWeyl}
\end{equation}
where $\chi = \pm$ is the chirality, $v$ the Fermi velocity, ${\bf p}$ the momentum, and ${\boldsymbol \sigma}$ are Pauli matrices. We consider the generic case of an extrinsic (doped) semimetal with positive chemical potential $\mu>0$ (the ``Weyl metal,'' although we continue referring to them as WSM). The spectrum of the Hamiltonian~\eqref{eq:HamiltonianWeyl} is linear with dispersion $\varepsilon_{\bf p} = \pm v |{\bf p}|$. Weyl nodes appear in pairs of opposite chirality~\cite{nielsen81a,nielsen81b,nielsen83}, and they can be separated by a wave vector ${\bf b}$ in the first Brillouin zone or by an energy offset $\hbar b_0$ in energy. The topological properties of a Weyl semimetal are manifested in the form of a $\theta$-term contribution to the action $S_\theta = \frac{e^2}{4\pi \hbar c} \int dt \int d^3r \, \theta \, {\bf E} \cdot {\bf B}$ with $\theta = 2 ({\bf b} \cdot {\bf r} - b_0 t)$~\cite{fujikawa79,zyuzin12,goswami13,hosur13}, where $e$ is the electron charge, ${\bf E}$ the electric field and ${\bf B}$ the magnetic field. If the bands are degenerate with $b_0 = {\bf b} = 0$ (the so-called Dirac semimetal), the system does not possess topological properties. The $\theta$ term changes the electromagnetic response of the material in the bulk medium by altering the constitutive relation that links the displacement field and the electric field~\cite{wilczek87,grushin12,hosur15,kargarian15,zyuzin15,pellegrino15,goswami15b}, which in frequency space reads
\begin{equation}
{\bf D} = \biggl(\varepsilon_\infty + \frac{4 \pi i}{\omega} \sigma\biggr) {\bf E} + \frac{i e^2}{\pi \hbar \omega} (\nabla \theta) \times {\bf E} + \frac{i e^2}{\pi \hbar c \omega} \dot{\theta} \, {\bf B} , \label{eq:defD}
\end{equation}
where $\varepsilon_\infty$ is the static dielectric constant of the medium and $\sigma$ the conductivity. The first term in parentheses is the standard term as in normal metals, and the last two terms arise due to the chiral anomaly. The gradient term in $\theta$ describes the contribution of an anomalous Hall current, and the last time-derivative term describes the chiral magnetic effect~\cite{hosur13}. Weyl semimetals have recently been reported for TaAs~\cite{xu15,lv15}, NbAs~\cite{xu15b}, YbMnBi${}_2$~\cite{borisenko15}, and Eu${}_2$Ir${}_2$O${}_7$~\cite{sushkov15}. In addition, semimetals with degenerate bands (Dirac semimetals) are reported for Cd${}_3$As${}_2$~\cite{borisenko14,liang15,neupane14}, ZrTe${}_5$~\cite{li14}, and Na${}_3$Bi~\cite{liu14}. They are parent materials from which nontrivial topological behavior is induced by symmetry breaking, for example, by applying an external magnetic field~\cite{kim13,li14,xiong15,goswami15}. These experimental results, which do not necessarily have an exclusive interpretation in terms of the chiral anomaly~\cite{goswami15}, motivate the search for direct signatures of Weyl semimetals that are of topological origin, i.e., effects that are not explained by the linear semimetallic Dirac dispersion and hence are not found in Dirac semimetals or small-gap semiconductors. We establish that SPP carry distinctive observable features in Weyl systems arising purely from their topological properties.

%++++++++++++++++++++++++++++++++++++++++++
\begin{figure*}[t!]
\begin{center}
\scalebox{0.69}{\includegraphics{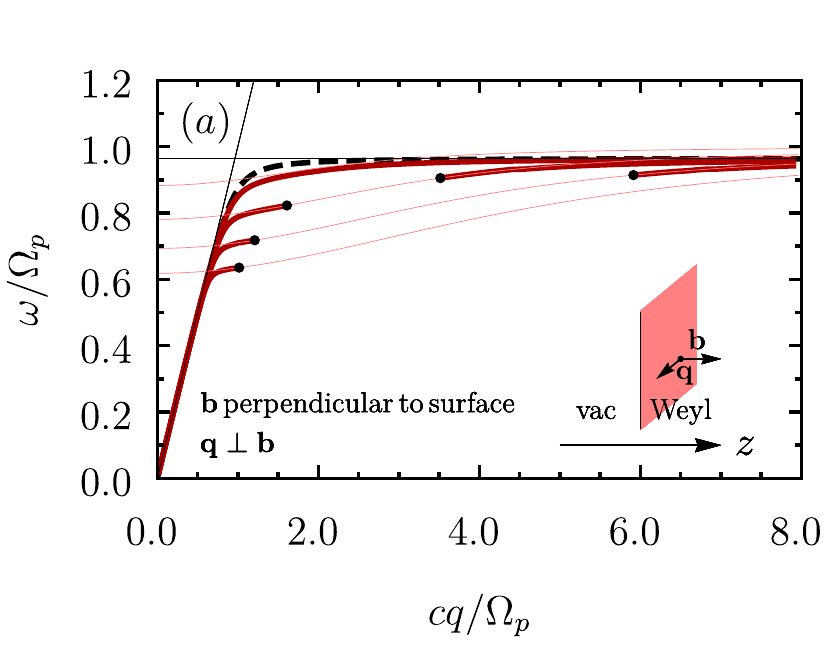}}
\scalebox{0.69}{\includegraphics{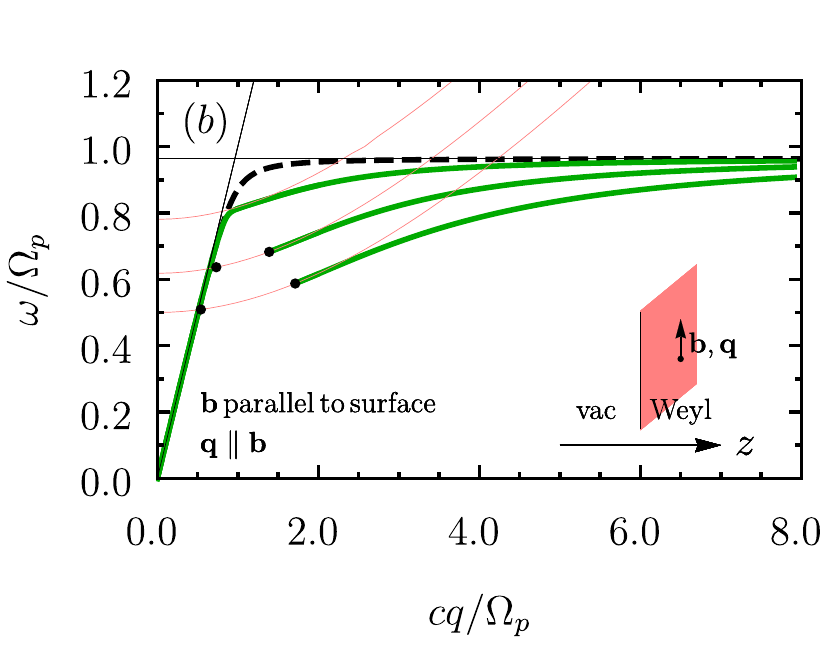}}
\scalebox{0.69}{\includegraphics{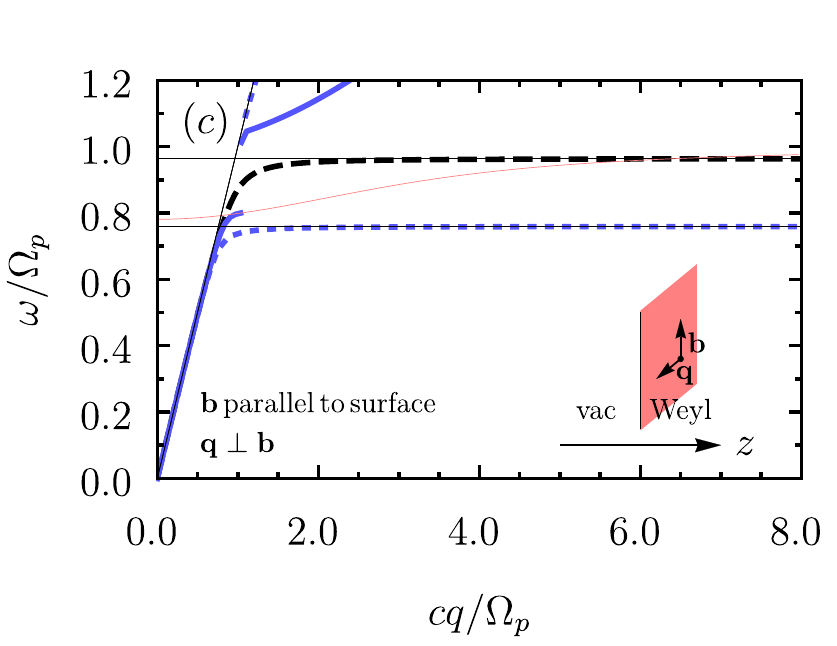}}
\caption{Surface plasmon dispersion of a Weyl semimetal with broken time-reversal symmetry for different values of (top to bottom) (a) $\omega_b/\Omega_p=0.25,0.5,0.75$, and $1$; (b) $0.5,1$, and $1.5$; and (c) $0.5$. In (c), the continuous blue line denotes positive wave numbers $q>0$ and the dotted blue line $q<0$. In all plots, the bulk plasmon dispersion is indicated by thin red lines. The thin black lines denote the asymptotic light line and the nonretarded frequency, respectively, and the black dashed line indicates the SPP dispersion of a standard Dirac material. As discussed in the main text, the black dots mark the points where the SPP hybridizes with the bulk plasmon mode and is damped.}
\label{fig:SPP1}
\end{center}
\end{figure*}
%++++++++++++++++++++++++++++++++++++++++++

Here, we show that the topological properties of Weyl semimetals affect the surface plasmon polariton dispersion. There are two main results of this work: First, because a WSM is an optically anisotropic medium, the surface plasmon dispersion depends on the Weyl node separation. The effect is strongest in the retarded limit (i.e., small wave vector), where the magnitude of the wave vector is comparable to the bulk plasmon frequency (divided by $c$, the velocity of light). We find that for a time-reversal broken WSM {\it without} external magnetic field, the SPP dispersion resembles retarded {\it magneto}plasmon modes in standard metals. In particular, for certain orientations of the surface, we predict a nonreciprocal dispersion, i.e., a dispersion that depends on the sign of the wave vector. As the second main result of this work, we predict that the Weyl surface magnetoplasmon modes possess an anomalous longitudinal magnetic field dependence that is absent for standard metals. This effect is caused by the anomalous magnetic field dependence of the WSM longitudinal conductivity (``negative magnetoresistance'')~\cite{son13,pellegrino15,spivak15}, a direct consequence of the chiral anomaly. In the remainder of this Rapid Communication, we derive both effects and discuss their experimental implications. All the algebraic details are provided in the Supplemental Material.

Surface plasmon polaritons are solutions of Maxwell's equations localized at the interface of two media. We consider the following geometry: A Weyl semimetal fills the positive half volume $z>0$, and a vacuum for $z<0$. The WSM-vacuum interface lies in the $x y$ plane. For simplicity, we restrict the analysis to a single pair of Weyl nodes (although our results are obviously valid for WSM with arbitrary pairs of nodes). Since we have translational invariance along the interface, the SSP are parametrized by the parallel wave vector ${\bf q} = (q_x, q_y)$. We search for electric fields of the form
\begin{equation}
{\bf E}^j = (E_x^j,E_y^j,E_z^j) e^{i q_x x + i q_y y} e^{- i \omega t} e^{- \kappa_j |z|} , \label{eq:ansatz}
\end{equation}
which decay exponentially away from the boundary, i.e., for which ${\rm Re} \, \kappa > 0$, and we label $j=0$ on the vacuum side and $j>0$ enumerates the solutions in the WSM. The decay constants $\kappa_j$ are determined from a solution of the wave equation
\begin{equation}
\nabla \times (\nabla \times {\bf E}) = - \frac{1}{c^2} \frac{\partial^2}{\partial t^2} {\bf D} , \label{eq:waveequation}
\end{equation}
where on the vacuum side, we have ${\bf D} = {\bf E}$, and for the WSM, ${\bf D}$ is given by Eq.~\eqref{eq:defD} (the magnetic field in this expression is related to the electric field by Faraday's law ${\bf B} = \frac{c}{i \omega} \nabla \times {\bf E}$). Substituting the ansatz~\eqref{eq:ansatz} in Eq.~\eqref{eq:waveequation}, we obtain a linear system of equations. The zeros of the determinant of this system yield $\kappa_j$. In general, on the WSM side, it turns out that there are two solutions of Eq.~\eqref{eq:waveequation} with exponentially decaying field. We demand the continuity of the parallel components of electric and magnetic fields ($E_{x/y}^1 = E_{x/y}^2$ and $B_{x/y}^1 = B_{x/y}^2$) and of the perpendicular components of the displacement fields ($D_z^1=D_z^2$ and $B_z^1 = B_z^2$). This gives four linearly independent conditions that determine the surface plasmon dispersion as well as the relative magnitude of the fields.

In the following, we assume that the dielectric tensor does not depend on the wavelength or the position inside the WSM. This approximation applies if the inverse wave vector of the SPP is large compared to the Thomas-Fermi length, which in a WSM is proportional to the inverse Fermi wave vector~\cite{dassarma15,li15}. In this case, the diagonal component of the dielectric tensor is $\varepsilon_1(\omega) = \varepsilon_\infty (1 - \frac{\Omega_p^2}{\omega^2})$, where $\Omega_p^2 = \frac{4 \alpha}{3 \pi} (\frac{\mu}{\hbar})^2$ denotes the bulk plasmon frequency~\cite{lv13,hofmann15,zhou15} with $\alpha = e^2/\hbar v \varepsilon_\infty$ being the finestructure constant of the WSM.

We first discuss the results for the SPP of a Dirac semimetal (for which ${\bf b} = {\bf 0}$ and $b_0 = 0$). The SPP solves
\begin{equation}
\varepsilon_1(\omega) + \kappa_0 = 0 ,
\end{equation}
with $\kappa_0 = \sqrt{q^2-\omega^2/c^2}$. This coincides with the conventional SPP condition in standard metals~\cite{ritchie57,pitarke07}. The surface plasmon dispersion is indicated by a thick black dashed line in Figs.~\ref{fig:SPP1}~(a)-(c). In the fully retarded limit $cq \ll \Omega_p$, the SPP follows the light-line $\omega = cq$ [thin black line in Figs.~\ref{fig:SPP1}(a)-\ref{fig:SPP1}(c)] and turns over in the hydrodynamic limit $cq \gg\Omega_p$ to a constant value $\omega = \sqrt{\varepsilon_\infty/(\varepsilon_\infty + 1)} \Omega_p$ (horizontal thin black line), which solves $\varepsilon_1(\omega) = - 1$. In particular, for $\varepsilon_\infty = 1$, this coincides with the famous result by Ritchie, $\omega = \Omega_p/\sqrt{2}$~\cite{ritchie57}. However, these surface plasmon modes are different from ordinary metals since they are purely quantum with $\hbar$ appearing explicitly~\cite{dassarma09}: $\Omega_p^2 \sim \alpha n^{2/3} \sim n^{2/3}/\hbar$. Furthermore, they show a sub-linear density dependence as opposed to a linear density-dependence for $\Omega_p^2$ in ordinary metals. Electron interactions can introduce a logarithmic correction to this scaling through charge renormalization~\cite{hofmann15}. We predict that the linear dispersion of a Dirac semimetal is manifested in a nonlinear dependence of the squared SPP mode frequency on doping density. Note that the characteristic scaling behavior may not only be probed by varying the doping density but also by finite-temperature measurements~\cite{sushkov15,hofmann15}.

We now present results for WSMs with broken time-reversal symmetry (${\bf b} \neq {\bf 0}$) and broken parity ($b_0 \neq 0$) for $\varepsilon_\infty = 13$ as measured in Eu${}_2$Ir${}_2$O${}_7$~\cite{sushkov15}. The results are shown in Fig.~\ref{fig:SPP1} for three relevant configurations: (a) ${\bf b}$ perpendicular to the sample surface, where the surface plasmon dispersion depends only on the magnitude of the parallel wave vector ${\bf q}$; (b) ${\bf b}$ parallel to the surface with ${\bf q}$ parallel to ${\bf b}$; and (c) ${\bf b}$ parallel to the surface with ${\bf q}$ perpendicular to ${\bf b}$. The chiral anomaly induces an off-diagonal term in the dielectric tensor $i \varepsilon_2(\omega) = i \varepsilon_\infty \omega_b/\omega$ with $\omega_b = 2 e^2 |{\bf b}|/\pi\hbar \varepsilon_\infty$. All the analytical calculational details are provided in the Supplemental Material. Figure~\ref{fig:SPP1}~(a) shows the SPP mode as a function of wave vector for four values of $\omega_b/\Omega_p = 0.5,1,1.5$, and $2$. The SPP deviates from the Dirac semimetal result (black dashed line) for intermediate wave vectors $cq \sim \Omega_p$ and departs from the light line at smaller wave number and energy. For comparison, we include as thin red lines the corresponding bulk plasmon modes for which one of the decay constants vanishes $\kappa=0$ and the plasmon is no longer confined to the surface. As is evident from the plots, for some wave vectors, bulk and surface modes are degenerate, while in other regions (marked by black end points), the SPP vanishes. Here, a generalized SPP still exists, but with a complex wave vector ${\bf q}$, indicating a coupling of surface and bulk modes~\cite{wallis74}. It is interesting to note that the geometry [Fig.~\ref{fig:SPP1}(a)] shows signs of the chiral anomaly, even though another characteristic signature of WSM -- topological Fermi arc surface states -- are absent in this configuration. Similar features are seen in Fig.~\ref{fig:SPP1}(b), which is shown for three different values $\omega_b/\Omega_p = 0.5,1$, and $1.5$. Both cases shown in Figs.~\ref{fig:SPP1}(a) and \ref{fig:SPP1}(b) are reciprocal, i.e., the dispersion is independent of the sign of $q$. In case (c), however, the SPP dispersion is nonreciprocal. For positive $q>0$ (blue dotted line), there is a transition from the light line to an asymptotic nonretarded constant frequency. For negative $q<0$, the dispersion has a discontinuity as it merges with the bulk plasmon mode, at which point it jumps to a higher frequency. In particular, there exists a frequency range where the system supports only modes with $q<0$. The nonreciprocity that we report could have interesting technological applications~\cite{camley87}. While nonreciprocal SSP in normal metals require magnetic fields or impurities~\cite{hartstein74}, nonreciprocity is a fundamental intrinsic material property of a WSM arising from its topological nature.

Strikingly, the SPP with ${\bf b} \neq {\bf 0}$ resemble, on a qualitative level, SPP of an ordinary metal in the presence of an external magnetic field~\cite{chiu72,wallis74,kushwaha01}, even though they have quite a different origin. Hence, the topological contribution to the dielectric tensor, which stems from an anomalous Hall displacement current, induces an ``anomalous surface magnetoplasmon.'' This is a central result of our work.

\begin{figure}[t!]
\begin{center}
\scalebox{0.69}{\includegraphics{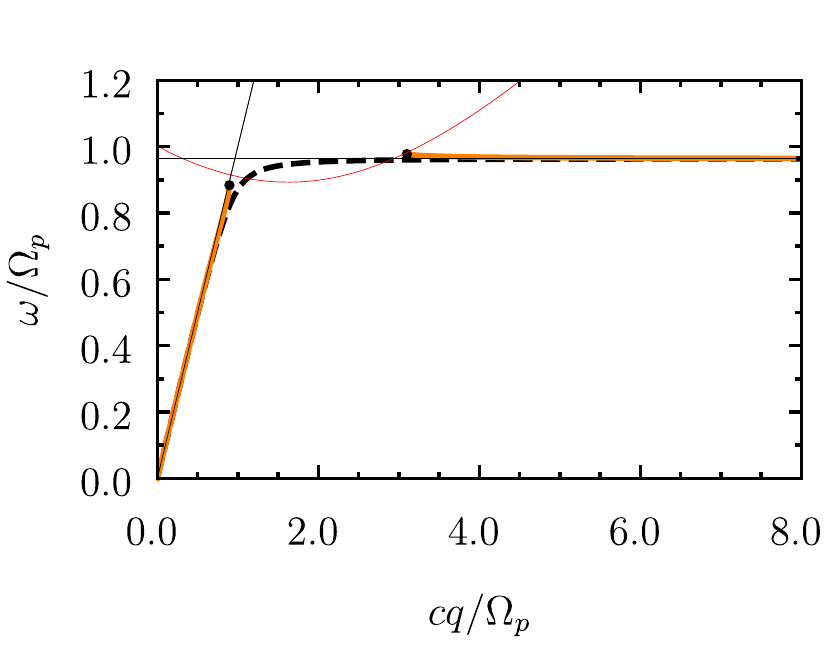}}
\caption{Surface plasmon dispersion of a Weyl semimetal with broken inversion symmetry for $\omega_{b_0}/\Omega_p = 0.5$, and $1$. The notation is the same as in Fig.~\ref{fig:SPP1}.}
\label{fig:SPP2}
\end{center}
\end{figure}

The retarded SPP dispersion of a WSM with $b_0 \neq 0$ (and ${\bf b} = {\bf 0}$) is shown in Fig.~\ref{fig:SPP2} for $\omega_{b_0}/\Omega_p = 0.5$. The dispersion solves
\begin{align}
&(\kappa_0 + \kappa_{1b}) \left[\kappa_{1a} \frac{\omega^2}{c^2} + \kappa_0 (q^2 - \kappa_{1a}^2)\right]  \nonumber \\
   & \times \left(\kappa_{1b}^2 - q^2 + \frac{\omega^2}{c^2} \varepsilon_1\right) + (\kappa_{1a} \leftrightarrow \kappa_{1b})  = 0 , \label{eq:b0SPPcondition}
  \end{align}
  where the decay constants are $\kappa_{1a/b} = q^2-\frac{1}{2} \frac{\omega^4}{c^4} \varepsilon _2^2-\frac{\omega^2}{c^2} \varepsilon _1 \pm \frac{1}{c^4} \sqrt{\frac{1}{4} \omega ^8 \varepsilon _2^4+c^2 \omega ^6 \varepsilon _1 \varepsilon _2^2}$ and $\varepsilon_2 = 2 e^2 b_0 \Omega_p/\pi \hbar c \omega^2 = \varepsilon_\infty \omega_{b_0}^2/\omega^2$, with $\omega_{b_0}^2 = 2 e^2 b_0 \Omega_p/\pi \hbar c \varepsilon_\infty$.  There is no dependence on the direction of the parallel wave number ${\bf q}$. In addition to the changed SPP dispersion, another observable effect would be a tilt of the field polarization out of the sagittal plane, as suggested for 3D topological insulators~\cite{karch11}.

We now consider WSM surface magnetoplasmon modes, which turn out to have an unusual magnetic field dependence. As this effect is distinct from the zero-field anomalous SPP discussed so far, we restrict our attention to the nonretarded limit and neglect corrections due to the separation of the Weyl nodes. The dielectric tensor takes the form $\varepsilon_1(\omega) = \varepsilon_\infty + \frac{4\pi i}{\omega} \sigma$. The conductivity can be derived in a semiclassical framework, in which the Berry curvature modifies the semiclassical equation of motion~\cite{xiao10,son13,lundgren14,pellegrino15,spivak15}. In particular, the Berry curvature induces a longitudinal magnetoconductivity~\cite{son13,spivak15}
\begin{equation}
\sigma_\parallel(\omega) = \frac{i \Omega_p^2}{4 \pi \omega} \biggl(1 + \frac{\alpha}{\pi} \frac{\omega_c^2}{\Omega_p^2}\biggr) , \label{eq:sigmaparallel}
\end{equation}
whereas in an ordinary metal, there is only the first Drude term and no dependence on the magnetic field. Here, $\omega_c = e v^2 B/\mu c$ denotes the cyclotron frequency and Eq.~\eqref{eq:sigmaparallel} holds for frequencies $\omega\gg \tau^{-1}$, where $\tau$ is the inelastic intranode scattering time~\cite{son13}. Note that the relative strength of the Drude and the anomalous term depends on the magnetic field and the doping density. The remaining components of the conductivity tensor take the standard Drude form $\sigma_\perp = i \Omega_p^2 \omega/4\pi(\omega^2 - \omega_c^2)$ and $\sigma_{xy} = \Omega_p^2 \omega_c/4 \pi(\omega^2-\omega_c^2)$, provided that Berry curvature corrections to the density of states and the intrinsic orbital moment are neglected~\cite{pellegrino15}. In the nonretarded limit, the electric field is given by ${\bf E} = - \nabla \phi$, where the electrostatic potential solves Poisson's equation $\nabla^2 \phi = 4 \pi \rho$ and $\rho$ is the induced charge density. On the vacuum side, the potential solves $\nabla^2 \phi = 0$. Combining Poisson's equation with the current and the continuity equation, we obtain (setting $\varepsilon_\infty = 1$)
\begin{equation}
 \nabla^2 \phi + \frac{4\pi i}{\omega} \nabla \cdot (\sigma \nabla \phi) = 0 . \label{eq:poisson}
\end{equation}
We make the ansatz $\phi = \phi_i e^{i q_x x + i q_y y} e^{-i \omega t} e^{- \kappa_j |z|}$ and impose the continuity of $\phi$ at the boundary. Integrating Eq.~\eqref{eq:poisson} across the interface, we find the second boundary condition
\begin{equation}
 \phi'(0+) - \phi'(0-) + \frac{4\pi i}{\omega} [\sigma \nabla \phi]_z(0+) = 0 , \label{eq:spboundary}
\end{equation}
where the prime denotes a derivative with respect to $z$. These conditions are of course equivalent to demanding the continuity of the parallel components of ${\bf E}$ and the perpendicular component of ${\bf D}$.

%++++++++++++++++++++++++++++++++++++++++++
%++++++++++++++++++++++++++++++++++++++++++

For a magnetic field parallel to the surface, we obtain the surface plasmon condition
\begin{align}
 \omega^2 - (\Omega_p^2 - \omega^2 + \omega_c^2) \Bigl(1+ \frac{\alpha}{\pi} \frac{\omega_c^2}{\Omega_p^2}\Bigr) \cos^2\theta \nonumber \\
 - 2 \omega \omega_c \sin \theta - (\Omega_p^2  - \omega^2) \sin^2\theta &= 0 ,
\end{align}
where $\theta$ is the relative angle between the magnetic field and the parallel momentum ${\bf q}$. The term proportional to $\sin \theta$ implies that the surface magnetoplasmon is nonreciprocal, i.e., the frequency depends on the sign of $q_y$. For a dispersion perpendicular to the magnetic field ($\cos \theta = 0$), we find $\omega^2 - \omega \omega_c {\rm sgn}(q) - \Omega_p^2/2 = 0$, which is the same form as in ordinary metals~\cite{fetter85}. There is a modification of the plasmon dispersion for modes that propagate along the magnetic field ($\sin \theta = 0$), for which we find
\begin{equation}
 \omega^2 = \frac{1 + \frac{\alpha}{\pi} \frac{\omega_c^2}{\Omega_p^2}}{2 + \frac{\alpha}{\pi} \frac{\omega_c^2}{\Omega_p^2}} \, \bigl(\Omega_p^2 + \omega_c^2\bigr) . \label{eq:plasmonparallel}
\end{equation}
The anomalous correction to $\sigma_\parallel$ changes the magnetic field dependence of the plasmon frequency. Comparing with the standard surface plasmon relation $\omega^2 = \varepsilon_\infty/(\varepsilon_\infty + 1) \Omega_p^2$, we interpret the correction as an anomalous contribution to the dielectric constant of the medium $\Delta \varepsilon_\infty = \alpha \omega_c^2/\pi \Omega_p^2 \sim B^2 n^{-4/3}$, which acquires a magnetic field dependence. If the anomalous term dominates (at small $\alpha$ and doping), the plasmon mode is equal to the bulk plasmon frequency, $\omega^2 = \Omega_p^2$. For a magnetic field that is perpendicular to the surface, we obtain the same anomalous surface plasmon dispersion as in Eq.~\eqref{eq:plasmonparallel}.

In summary, we predict a rich (and experimentally observable) structure of surface plasmon polaritons in doped Weyl semimetals. In particular, we show the following: (a) There is a quantum surface plasmon mode with unusual density dependence; (b) for broken inversion symmetry, the retarded dispersion is strongly affected by the chemical potential imbalance; (c) for broken time-reversal symmetry, the dispersion depends on the Weyl node separation, which acts similar to an internal magnetic field; (d) a nonreciprocal dispersion arises naturally even without external magnetic fields; and (e) the magnetoplasmon mode acquires an additional longitudinal magnetic field dependence. These effects are experimentally observable signatures of the chiral anomaly in Weyl semimetals and could point the way to future technological applications of these systems.

\begin{acknowledgments}
This work is supported by LPS-MPO-CMTC, Microsoft Q, and JQI-NSF-PFC (J.H. and S.D.S.), and Gonville and Caius College (J.H.).
\end{acknowledgments}


%%%%%%%%% Supplemental material %%%%%%%%%%
\widetext
\pagebreak

\begin{center}
\textbf{\large Supplemental Material: Surface plasmon polaritons in topological Weyl semimetals}
\end{center}

\begin{center}
Johannes Hofmann${}^{1,2}$ and Sankar Das Sarma${}^1$

\it{${}^1$Condensed Matter Theory Center and Joint Quantum Institute, Department of Physics, University of Maryland, College Park, Maryland 20742-4111 USA}

\it{${}^2$T.C.M. Group, Cavendish Laboratory, University of Cambridge, Cambridge CB3 0HE, United Kingdom}

\end{center}

\setcounter{equation}{0}
\setcounter{figure}{0}
\setcounter{table}{0}
\setcounter{page}{1}
\makeatletter
\renewcommand{\theequation}{S\arabic{equation}}
\renewcommand{\thefigure}{S\arabic{figure}}
\renewcommand{\bibnumfmt}[1]{[S#1]}
\renewcommand{\citenumfont}[1]{S#1}

In this supplemental material, we provide details on the calculation of the retarded surface plasmon polariton dispersion. The interface between the vacuum and the Weyl semimetal (WSM) is in the $x$-$y$ plane at $z=0$, with the vacuum for $z<0$ and the WSM for $z>0$. The evanescent ansatz~\eqref{eq:ansatz} for the electric field at the boundary solves the wave equation~\eqref{eq:waveequation}, where the displacement field ${\bf D}$ is related to the electric field ${\bf E}$ through the dielectric tensor $\hat{\varepsilon}(\omega)$, Eq.~\eqref{eq:defD}. Hence, the electric field solves a homogeneous equation $\hat{M} {\bf E} = 0$, where
\begin{equation}
\hat{M} = \begin{pmatrix} 
q_y^2 - \kappa_j^2 & - q_x q_y & \mp i q_x \kappa_j \\
- q_x q_y & q_x^2 - \kappa_j^2 & \mp i q_y \kappa_j \\
\mp i q_x \kappa_j & \mp i q_y \kappa_j & q_x^2 + q_y^2
\end{pmatrix} - \frac{\omega^2}{c^2} \hat{\varepsilon}(\omega) ,
\end{equation}
with the positive sign on the vacuum side and negative on the WSM side. The zeros of the determinant of $\hat{M}$ determine the decay constant $\kappa_j$. On the vacuum side ($j=0$), we find $\kappa_0^2 = q^2 - \frac{\omega^2}{c^2}$ for any orientation of $q$. On the WSM side ($j=1$), we have for the various geometries shown in Fig.~\ref{fig:SPP1}: \\ \\
Case (a): ${\bf b} = (0,0,b), {\bf q} = (0,q,0)$
\begin{align}
\hat{M} =
\begin{pmatrix}
q^2 - \kappa_1^2 - \frac{\omega^2}{c^2} \varepsilon_1 & - \frac{i \omega^2}{c^2} \varepsilon_2 & 0 \\
\frac{i \omega^2}{c^2} \varepsilon_2(\omega) & - \kappa_1^2 - \frac{\omega^2}{c^2}\varepsilon_1 & - i q \kappa_1 \\
0 & - i q \kappa_1 & q^2 - \frac{\omega^2}{c^2} \varepsilon_1
\end{pmatrix} ,
\qquad
\kappa_1^2 &= q^2 - \frac{\omega^2}{c^2} \varepsilon_1 \pm \frac{1}{\varepsilon_1} \sqrt{- \frac{\omega^2}{c^2} \varepsilon_2^2 \Bigl(q^2 - \frac{\omega^2}{c^2} \varepsilon_1\Bigr)} \label{eq:supA}
\end{align}
Case (b): ${\bf b} = (b,0,0), {\bf q} = (q,0,0)$
\begin{align}
\hat{M} =
\begin{pmatrix}
- \kappa_1^2 - \frac{\omega^2}{c^2} \varepsilon_1& 0 & - i q \kappa_1 \\
0 & q^2 - \kappa_1^2 - \frac{\omega^2}{c^2} \varepsilon_1 & -\frac{i \omega^2}{c^2} \varepsilon_2 \\
- i q \kappa_1 & \frac{i \omega^2}{c^2} \varepsilon_2 & q^2 - \frac{\omega^2}{c^2} \varepsilon_1
\end{pmatrix} ,
\qquad
\kappa_1^2 &= q^2 + \frac{\omega^2}{c^2} \biggl( \frac{\varepsilon _2^2}{2 \varepsilon_1} - \varepsilon _1 \biggr) \pm \frac{1}{2 \varepsilon_1} \sqrt{4 q^2 \frac{\omega^2}{c^2} \varepsilon_1 \varepsilon_2^2 + \frac{\omega^4}{c^4} \varepsilon_2^4}
\end{align}
{\rm Case (c)}: ${\bf b} = (b,0,0), {\bf q} = (0,q,0)$
\begin{align}
\hat{M} =
\begin{pmatrix}
q^2 - \kappa_1^2 - \frac{\omega^2}{c^2} \varepsilon_1 & 0 & 0 \\
0 & - \kappa_1^2 - \frac{\omega^2}{c^2} \varepsilon_1 & - i q \kappa_1 - \frac{i \omega^2}{c^2} \varepsilon_2 \\
0 & - i q \kappa_1 + \frac{i \omega^2}{c^2} \varepsilon_2 & q^2 - \frac{\omega^2}{c^2} \varepsilon_1
\end{pmatrix} ,
\qquad
\kappa_1^2 &= q^2 + \frac{\omega^2}{c^2} \biggl(\frac{\varepsilon_2^2}{\varepsilon_1} - \varepsilon_1\biggr) , \label{eq:supC}
\end{align}
where we define as in the main text $\varepsilon_1(\omega) = \varepsilon_\infty (1 - \frac{\Omega_p^2}{\omega^2})$ and $\varepsilon_2(\omega) = \varepsilon_\infty \omega_b/\omega$ with $\omega_b = 2 e^2 b/\pi\hbar \varepsilon_\infty$. In Eqs.~\eqref{eq:supA}-\eqref{eq:supC}, we also note the results for the decay constant $\kappa_1$ on the WSM side. For $b_0 \neq 0$ and ${\bf q} = (q,0,0)$ as shown in Fig.~\ref{fig:SPP2}, we have
\begin{align}
\hat{M} =
\begin{pmatrix}
- \kappa_1^2 - \frac{\omega^2}{c^2} \varepsilon_1(\omega) & \frac{c\kappa_1}{\Omega_p} \varepsilon_2 & - i q \kappa_1 \\
- \frac{c\kappa_1}{\Omega_p} \varepsilon_2 & q^2 - \kappa_1^2 - \frac{\omega^2}{c^2} \varepsilon_1(\omega) & \frac{i cq}{\Omega_p} \varepsilon_2 \\
- \frac{i cq}{\Omega_p} \varepsilon_2 & \frac{i \omega^2}{c^2} \varepsilon_2(\omega) & q^2 - \frac{\omega^2}{c^2} \varepsilon_1(\omega)
\end{pmatrix} ,
\qquad
\kappa_1^2 &= q^2 - \frac{\omega^2}{c^2} \varepsilon_1 - \frac{\omega^4}{2 c^4} \varepsilon_2^2 + \frac{1}{2} \sqrt{\frac{\omega^8}{c^8} \varepsilon_2^4 + 4 \frac{\omega^6}{c^6} \varepsilon_1 \varepsilon_2^2} ,
\end{align}
where we define $\varepsilon_2(\omega) = \varepsilon_\infty \omega_{b_0}^2/\omega^2$ with $\omega_{b_0}^2 = 2 e^2 b_0 \Omega_p/\pi \hbar c \varepsilon_\infty$. There are, in general, two linearly independent solutions for ${\bf E}$ on the vacuum and on the WSM side which correspond to modes that are localized at the interface. We impose as a boundary condition the continuity of the parallel component of the electric field ${\bf E}$ and the perpendicular component of the displacement field $D_z$ as well as the continuity of the magnetic field ${\bf B} = \frac{c}{i\omega} \nabla \times {\bf E}$. This gives a set of four linearly independent constraints. Setting the determinant of this constraint matrix equal to zero, we find the SPP condition for the various cases: \\ \\
Case (a):
\begin{equation}
 \epsilon_1 \kappa_0 \kappa_{1a} \kappa_{1b} (\kappa_{1a} + \kappa_{1b}) + q^2 \bigl[\kappa_{1a} \kappa_{1b} + \kappa_0 (\kappa_{1a}+\kappa_{1b}) + (\kappa_{1a}^2 + \kappa_{1a} \kappa_{1b} + \kappa_{1b}^2) \epsilon_1\bigr] - \frac{\omega^2}{c^2} \epsilon_1 \biggl((\kappa_{1a}+\kappa_{1b}) (\kappa_0 + \kappa_{1a} + \kappa_{1b} + q^2 (1 - \epsilon_1)\biggr) = 0 .
\end{equation}
Case (b):
\begin{equation}
\kappa_{1a} \kappa_{1b} (\kappa_0 + \kappa_{1a} + \kappa_{1b}) + \kappa_0 (\kappa_{1a}^2 + \kappa_{1a} \kappa_{1b} + \kappa_{1b}^2) \epsilon_1  + q^2 \bigl[\kappa_0 (1-\epsilon_1)+(\kappa_{1a} + \kappa_{1b}) \epsilon_1\bigr] - \frac{\omega^2}{c^2} \epsilon_1 \biggl(\kappa_{1a} + \kappa_{1b} + \kappa_0 (1 - \epsilon_1)\biggr)= 0 .
\end{equation}
Case (c):
\begin{equation}
\varepsilon_1 \kappa_1 + \kappa_0 (\varepsilon_1^2-\varepsilon_2^2) - q \varepsilon_2 = 0 .
\end{equation}
Here, we denote by $\kappa_{1a}$ and $\kappa_{1b}$ the two solutions corresponding to $\pm$ in Eqs.~\eqref{eq:supA}-\eqref{eq:supC}. For $b_0 \neq 0$, the condition is stated in Eq.~\eqref{eq:b0SPPcondition} of the main text. In the case ${\bf b} \neq 0$, the result is similar as for the as for a metal in a constant external magnetic field~\cite{suppchiu72,suppwallis74}, and the SPP conditions agree when correcting for the difference in the dielectric tensor. The explicit form of the dielectric components in a WSM is, of course, different from the Drude form in an external magnetic field.


\begin{thebibliography}{57}%
\makeatletter
\providecommand \@ifxundefined [1]{%
 \@ifx{#1\undefined}
}%
\providecommand \@ifnum [1]{%
 \ifnum #1\expandafter \@firstoftwo
 \else \expandafter \@secondoftwo
 \fi
}%
\providecommand \@ifx [1]{%
 \ifx #1\expandafter \@firstoftwo
 \else \expandafter \@secondoftwo
 \fi
}%
\providecommand \natexlab [1]{#1}%
\providecommand \enquote  [1]{``#1''}%
\providecommand \bibnamefont  [1]{#1}%
\providecommand \bibfnamefont [1]{#1}%
\providecommand \citenamefont [1]{#1}%
\providecommand \href@noop [0]{\@secondoftwo}%
\providecommand \href [0]{\begingroup \@sanitize@url \@href}%
\providecommand \@href[1]{\@@startlink{#1}\@@href}%
\providecommand \@@href[1]{\endgroup#1\@@endlink}%
\providecommand \@sanitize@url [0]{\catcode `\\12\catcode `\$12\catcode
  `\&12\catcode `\#12\catcode `\^12\catcode `\_12\catcode `\%12\relax}%
\providecommand \@@startlink[1]{}%
\providecommand \@@endlink[0]{}%
\providecommand \url  [0]{\begingroup\@sanitize@url \@url }%
\providecommand \@url [1]{\endgroup\@href {#1}{\urlprefix }}%
\providecommand \urlprefix  [0]{URL }%
\providecommand \Eprint [0]{\href }%
\providecommand \doibase [0]{http://dx.doi.org/}%
\providecommand \selectlanguage [0]{\@gobble}%
\providecommand \bibinfo  [0]{\@secondoftwo}%
\providecommand \bibfield  [0]{\@secondoftwo}%
\providecommand \translation [1]{[#1]}%
\providecommand \BibitemOpen [0]{}%
\providecommand \bibitemStop [0]{}%
\providecommand \bibitemNoStop [0]{.\EOS\space}%
\providecommand \EOS [0]{\spacefactor3000\relax}%
\providecommand \BibitemShut  [1]{\csname bibitem#1\endcsname}%
\let\auto@bib@innerbib\@empty
%</preamble>
\bibitem [{\citenamefont {Ritchie}(1957)}]{ritchie57}%
  \BibitemOpen
  \bibfield  {author} {\bibinfo {author} {\bibfnamefont {R.~H.}\ \bibnamefont
  {Ritchie}},\ }\href@noop {} {\bibfield  {journal} {\bibinfo  {journal} {Phys.
  Rev.}\ }\textbf {\bibinfo {volume} {106}},\ \bibinfo {pages} {874} (\bibinfo
  {year} {1957})}\BibitemShut {NoStop}%
\bibitem [{\citenamefont {Stern}\ and\ \citenamefont
  {Ferrell}(1960)}]{stern60}%
  \BibitemOpen
  \bibfield  {author} {\bibinfo {author} {\bibfnamefont {E.~A.}\ \bibnamefont
  {Stern}}\ and\ \bibinfo {author} {\bibfnamefont {R.~A.}\ \bibnamefont
  {Ferrell}},\ }\href@noop {} {\bibfield  {journal} {\bibinfo  {journal} {Phys.
  Rev.}\ }\textbf {\bibinfo {volume} {120}},\ \bibinfo {pages} {130} (\bibinfo
  {year} {1960})}\BibitemShut {NoStop}%
\bibitem [{\citenamefont {Powell}\ and\ \citenamefont {Swan}(1959)}]{powell59}%
  \BibitemOpen
  \bibfield  {author} {\bibinfo {author} {\bibfnamefont {C.~J.}\ \bibnamefont
  {Powell}}\ and\ \bibinfo {author} {\bibfnamefont {J.~B.}\ \bibnamefont
  {Swan}},\ }\href@noop {} {\bibfield  {journal} {\bibinfo  {journal} {Phys.
  Rev.}\ }\textbf {\bibinfo {volume} {115}},\ \bibinfo {pages} {869} (\bibinfo
  {year} {1959})}\BibitemShut {NoStop}%
\bibitem [{\citenamefont {Garc\'{\i}a~de Abajo}(2010)}]{garcia10}%
  \BibitemOpen
  \bibfield  {author} {\bibinfo {author} {\bibfnamefont {F.~J.}\ \bibnamefont
  {Garc\'{\i}a~de Abajo}},\ }\href@noop {} {\bibfield  {journal} {\bibinfo
  {journal} {Rev. Mod. Phys.}\ }\textbf {\bibinfo {volume} {82}},\ \bibinfo
  {pages} {209} (\bibinfo {year} {2010})}\BibitemShut {NoStop}%
\bibitem [{\citenamefont {Ritchie}\ \emph {et~al.}(1968)\citenamefont
  {Ritchie}, \citenamefont {Arakawa}, \citenamefont {Cowan},\ and\
  \citenamefont {Hamm}}]{ritchie68}%
  \BibitemOpen
  \bibfield  {author} {\bibinfo {author} {\bibfnamefont {R.~H.}\ \bibnamefont
  {Ritchie}}, \bibinfo {author} {\bibfnamefont {E.~T.}\ \bibnamefont
  {Arakawa}}, \bibinfo {author} {\bibfnamefont {J.~J.}\ \bibnamefont {Cowan}},
  \ and\ \bibinfo {author} {\bibfnamefont {R.~N.}\ \bibnamefont {Hamm}},\
  }\href@noop {} {\bibfield  {journal} {\bibinfo  {journal} {Phys. Rev. Lett.}\
  }\textbf {\bibinfo {volume} {21}},\ \bibinfo {pages} {1530} (\bibinfo {year}
  {1968})}\BibitemShut {NoStop}%
\bibitem [{\citenamefont {Marschall}\ \emph {et~al.}(1971)\citenamefont
  {Marschall}, \citenamefont {Fischer},\ and\ \citenamefont
  {Queisser}}]{marschall71}%
  \BibitemOpen
  \bibfield  {author} {\bibinfo {author} {\bibfnamefont {N.}~\bibnamefont
  {Marschall}}, \bibinfo {author} {\bibfnamefont {B.}~\bibnamefont {Fischer}},
  \ and\ \bibinfo {author} {\bibfnamefont {H.~J.}\ \bibnamefont {Queisser}},\
  }\href@noop {} {\bibfield  {journal} {\bibinfo  {journal} {Phys. Rev. Lett.}\
  }\textbf {\bibinfo {volume} {27}},\ \bibinfo {pages} {95} (\bibinfo {year}
  {1971})}\BibitemShut {NoStop}%
\bibitem [{\citenamefont {Rothenhausler}\ and\ \citenamefont
  {Knoll}(1988)}]{rothenhausler88}%
  \BibitemOpen
  \bibfield  {author} {\bibinfo {author} {\bibfnamefont {B.}~\bibnamefont
  {Rothenhausler}}\ and\ \bibinfo {author} {\bibfnamefont {W.}~\bibnamefont
  {Knoll}},\ }\href@noop {} {\bibfield  {journal} {\bibinfo  {journal}
  {Nature}\ }\textbf {\bibinfo {volume} {332}},\ \bibinfo {pages} {615}
  (\bibinfo {year} {1988})}\BibitemShut {NoStop}%
\bibitem [{\citenamefont {Malmqvist}(1993)}]{malmqvist93}%
  \BibitemOpen
  \bibfield  {author} {\bibinfo {author} {\bibfnamefont {M.}~\bibnamefont
  {Malmqvist}},\ }\href@noop {} {\bibfield  {journal} {\bibinfo  {journal}
  {Nature}\ }\textbf {\bibinfo {volume} {361}},\ \bibinfo {pages} {186}
  (\bibinfo {year} {1993})}\BibitemShut {NoStop}%
\bibitem [{\citenamefont {Srituravanich}\ \emph {et~al.}(2004)\citenamefont
  {Srituravanich}, \citenamefont {Fang}, \citenamefont {Sun}, \citenamefont
  {Luo},\ and\ \citenamefont {Zhang}}]{srituravanich04}%
  \BibitemOpen
  \bibfield  {author} {\bibinfo {author} {\bibfnamefont {W.}~\bibnamefont
  {Srituravanich}}, \bibinfo {author} {\bibfnamefont {N.}~\bibnamefont {Fang}},
  \bibinfo {author} {\bibfnamefont {C.}~\bibnamefont {Sun}}, \bibinfo {author}
  {\bibfnamefont {Q.}~\bibnamefont {Luo}}, \ and\ \bibinfo {author}
  {\bibfnamefont {X.}~\bibnamefont {Zhang}},\ }\href@noop {} {\bibfield
  {journal} {\bibinfo  {journal} {Nano Letters}\ }\textbf {\bibinfo {volume}
  {4}},\ \bibinfo {pages} {1085} (\bibinfo {year} {2004})}\BibitemShut
  {NoStop}%
\bibitem [{\citenamefont {Barnes}\ \emph {et~al.}(2003)\citenamefont {Barnes},
  \citenamefont {Dereux},\ and\ \citenamefont {Ebbesen}}]{barnes03}%
  \BibitemOpen
  \bibfield  {author} {\bibinfo {author} {\bibfnamefont {W.~L.}\ \bibnamefont
  {Barnes}}, \bibinfo {author} {\bibfnamefont {A.}~\bibnamefont {Dereux}}, \
  and\ \bibinfo {author} {\bibfnamefont {T.~W.}\ \bibnamefont {Ebbesen}},\
  }\href@noop {} {\bibfield  {journal} {\bibinfo  {journal} {Nature}\ }\textbf
  {\bibinfo {volume} {424}},\ \bibinfo {pages} {824} (\bibinfo {year}
  {2003})}\BibitemShut {NoStop}%
\bibitem [{\citenamefont {Vafek}\ and\ \citenamefont
  {Vishwanath}(2014)}]{vafek14}%
  \BibitemOpen
  \bibfield  {author} {\bibinfo {author} {\bibfnamefont {O.}~\bibnamefont
  {Vafek}}\ and\ \bibinfo {author} {\bibfnamefont {A.}~\bibnamefont
  {Vishwanath}},\ }\href@noop {} {\bibfield  {journal} {\bibinfo  {journal}
  {Annual Review of Condensed Matter Physics}\ }\textbf {\bibinfo {volume}
  {5}},\ \bibinfo {pages} {83} (\bibinfo {year} {2014})}\BibitemShut {NoStop}%
\bibitem [{\citenamefont {Wehling}\ \emph {et~al.}(2014)\citenamefont
  {Wehling}, \citenamefont {Black-Schaffer},\ and\ \citenamefont
  {Balatsky}}]{wehling14}%
  \BibitemOpen
  \bibfield  {author} {\bibinfo {author} {\bibfnamefont {T.}~\bibnamefont
  {Wehling}}, \bibinfo {author} {\bibfnamefont {A.}~\bibnamefont
  {Black-Schaffer}}, \ and\ \bibinfo {author} {\bibfnamefont {A.}~\bibnamefont
  {Balatsky}},\ }\href@noop {} {\bibfield  {journal} {\bibinfo  {journal}
  {Advances in Physics}\ }\textbf {\bibinfo {volume} {63}},\ \bibinfo {pages}
  {1} (\bibinfo {year} {2014})}\BibitemShut {NoStop}%
\bibitem [{\citenamefont {Nielsen}\ and\ \citenamefont
  {Ninomiya}(1981{\natexlab{a}})}]{nielsen81a}%
  \BibitemOpen
  \bibfield  {author} {\bibinfo {author} {\bibfnamefont {H.}~\bibnamefont
  {Nielsen}}\ and\ \bibinfo {author} {\bibfnamefont {M.}~\bibnamefont
  {Ninomiya}},\ }\href@noop {} {\bibfield  {journal} {\bibinfo  {journal}
  {Nuclear Physics B}\ }\textbf {\bibinfo {volume} {185}},\ \bibinfo {pages}
  {20} (\bibinfo {year} {1981}{\natexlab{a}})}\BibitemShut {NoStop}%
\bibitem [{\citenamefont {Nielsen}\ and\ \citenamefont
  {Ninomiya}(1981{\natexlab{b}})}]{nielsen81b}%
  \BibitemOpen
  \bibfield  {author} {\bibinfo {author} {\bibfnamefont {H.}~\bibnamefont
  {Nielsen}}\ and\ \bibinfo {author} {\bibfnamefont {M.}~\bibnamefont
  {Ninomiya}},\ }\href@noop {} {\bibfield  {journal} {\bibinfo  {journal}
  {Nuclear Physics B}\ }\textbf {\bibinfo {volume} {193}},\ \bibinfo {pages}
  {173} (\bibinfo {year} {1981}{\natexlab{b}})}\BibitemShut {NoStop}%
\bibitem [{\citenamefont {Nielsen}\ and\ \citenamefont
  {Ninomiya}(1983)}]{nielsen83}%
  \BibitemOpen
  \bibfield  {author} {\bibinfo {author} {\bibfnamefont {H.}~\bibnamefont
  {Nielsen}}\ and\ \bibinfo {author} {\bibfnamefont {M.}~\bibnamefont
  {Ninomiya}},\ }\href@noop {} {\bibfield  {journal} {\bibinfo  {journal}
  {Physics Letters B}\ }\textbf {\bibinfo {volume} {130}},\ \bibinfo {pages}
  {389} (\bibinfo {year} {1983})}\BibitemShut {NoStop}%
\bibitem [{\citenamefont {Fujikawa}(1979)}]{fujikawa79}%
  \BibitemOpen
  \bibfield  {author} {\bibinfo {author} {\bibfnamefont {K.}~\bibnamefont
  {Fujikawa}},\ }\href@noop {} {\bibfield  {journal} {\bibinfo  {journal}
  {Phys. Rev. Lett.}\ }\textbf {\bibinfo {volume} {42}},\ \bibinfo {pages}
  {1195} (\bibinfo {year} {1979})}\BibitemShut {NoStop}%
\bibitem [{\citenamefont {Zyuzin}\ and\ \citenamefont
  {Burkov}(2012)}]{zyuzin12}%
  \BibitemOpen
  \bibfield  {author} {\bibinfo {author} {\bibfnamefont {A.~A.}\ \bibnamefont
  {Zyuzin}}\ and\ \bibinfo {author} {\bibfnamefont {A.~A.}\ \bibnamefont
  {Burkov}},\ }\href@noop {} {\bibfield  {journal} {\bibinfo  {journal} {Phys.
  Rev. B}\ }\textbf {\bibinfo {volume} {86}},\ \bibinfo {pages} {115133}
  (\bibinfo {year} {2012})}\BibitemShut {NoStop}%
\bibitem [{\citenamefont {Goswami}\ and\ \citenamefont
  {Tewari}(2013)}]{goswami13}%
  \BibitemOpen
  \bibfield  {author} {\bibinfo {author} {\bibfnamefont {P.}~\bibnamefont
  {Goswami}}\ and\ \bibinfo {author} {\bibfnamefont {S.}~\bibnamefont
  {Tewari}},\ }\href@noop {} {\bibfield  {journal} {\bibinfo  {journal} {Phys.
  Rev. B}\ }\textbf {\bibinfo {volume} {88}},\ \bibinfo {pages} {245107}
  (\bibinfo {year} {2013})}\BibitemShut {NoStop}%
\bibitem [{\citenamefont {Hosur}\ and\ \citenamefont {Qi}(2013)}]{hosur13}%
  \BibitemOpen
  \bibfield  {author} {\bibinfo {author} {\bibfnamefont {P.}~\bibnamefont
  {Hosur}}\ and\ \bibinfo {author} {\bibfnamefont {X.}~\bibnamefont {Qi}},\
  }\href@noop {} {\bibfield  {journal} {\bibinfo  {journal} {Comptes Rendus
  Physique}\ }\textbf {\bibinfo {volume} {14}},\ \bibinfo {pages} {857}
  (\bibinfo {year} {2013})}\BibitemShut {NoStop}%
\bibitem [{\citenamefont {Wilczek}(1987)}]{wilczek87}%
  \BibitemOpen
  \bibfield  {author} {\bibinfo {author} {\bibfnamefont {F.}~\bibnamefont
  {Wilczek}},\ }\href@noop {} {\bibfield  {journal} {\bibinfo  {journal} {Phys.
  Rev. Lett.}\ }\textbf {\bibinfo {volume} {58}},\ \bibinfo {pages} {1799}
  (\bibinfo {year} {1987})}\BibitemShut {NoStop}%
\bibitem [{\citenamefont {Grushin}(2012)}]{grushin12}%
  \BibitemOpen
  \bibfield  {author} {\bibinfo {author} {\bibfnamefont {A.~G.}\ \bibnamefont
  {Grushin}},\ }\href@noop {} {\bibfield  {journal} {\bibinfo  {journal} {Phys.
  Rev. D}\ }\textbf {\bibinfo {volume} {86}},\ \bibinfo {pages} {045001}
  (\bibinfo {year} {2012})}\BibitemShut {NoStop}%
\bibitem [{\citenamefont {Hosur}\ and\ \citenamefont {Qi}(2015)}]{hosur15}%
  \BibitemOpen
  \bibfield  {author} {\bibinfo {author} {\bibfnamefont {P.}~\bibnamefont
  {Hosur}}\ and\ \bibinfo {author} {\bibfnamefont {X.-L.}\ \bibnamefont {Qi}},\
  }\href@noop {} {\bibfield  {journal} {\bibinfo  {journal} {Phys. Rev. B}\
  }\textbf {\bibinfo {volume} {91}},\ \bibinfo {pages} {081106} (\bibinfo
  {year} {2015})}\BibitemShut {NoStop}%
\bibitem [{\citenamefont {Kargarian}\ \emph {et~al.}(2015)\citenamefont
  {Kargarian}, \citenamefont {Randeria},\ and\ \citenamefont
  {Trivedi}}]{kargarian15}%
  \BibitemOpen
  \bibfield  {author} {\bibinfo {author} {\bibfnamefont {M.}~\bibnamefont
  {Kargarian}}, \bibinfo {author} {\bibfnamefont {M.}~\bibnamefont {Randeria}},
  \ and\ \bibinfo {author} {\bibfnamefont {N.}~\bibnamefont {Trivedi}},\
  }\href@noop {} {\bibfield  {journal} {\bibinfo  {journal} {Scientific
  Reports}\ }\textbf {\bibinfo {volume} {5}},\ \bibinfo {pages} {12683 EP }
  (\bibinfo {year} {2015})}\BibitemShut {NoStop}%
\bibitem [{\citenamefont {Zyuzin}\ and\ \citenamefont
  {Zyuzin}(2015)}]{zyuzin15}%
  \BibitemOpen
  \bibfield  {author} {\bibinfo {author} {\bibfnamefont {A.~A.}\ \bibnamefont
  {Zyuzin}}\ and\ \bibinfo {author} {\bibfnamefont {V.~A.}\ \bibnamefont
  {Zyuzin}},\ }\href@noop {} {\bibfield  {journal} {\bibinfo  {journal} {Phys.
  Rev. B}\ }\textbf {\bibinfo {volume} {92}},\ \bibinfo {pages} {115310}
  (\bibinfo {year} {2015})}\BibitemShut {NoStop}%
\bibitem [{\citenamefont {Pellegrino}\ \emph {et~al.}(2015)\citenamefont
  {Pellegrino}, \citenamefont {Katsnelson},\ and\ \citenamefont
  {Polini}}]{pellegrino15}%
  \BibitemOpen
  \bibfield  {author} {\bibinfo {author} {\bibfnamefont {F.~M.~D.}\
  \bibnamefont {Pellegrino}}, \bibinfo {author} {\bibfnamefont {M.~I.}\
  \bibnamefont {Katsnelson}}, \ and\ \bibinfo {author} {\bibfnamefont
  {M.}~\bibnamefont {Polini}},\ }\href@noop {} {\bibfield  {journal} {\bibinfo
  {journal} {Phys. Rev. B}\ }\textbf {\bibinfo {volume} {92}},\ \bibinfo
  {pages} {201407} (\bibinfo {year} {2015})}\BibitemShut {NoStop}%
\bibitem [{\citenamefont {Goswami}\ \emph
  {et~al.}(2015{\natexlab{a}})\citenamefont {Goswami}, \citenamefont {Sharma},\
  and\ \citenamefont {Tewari}}]{goswami15b}%
  \BibitemOpen
  \bibfield  {author} {\bibinfo {author} {\bibfnamefont {P.}~\bibnamefont
  {Goswami}}, \bibinfo {author} {\bibfnamefont {G.}~\bibnamefont {Sharma}}, \
  and\ \bibinfo {author} {\bibfnamefont {S.}~\bibnamefont {Tewari}},\
  }\href@noop {} {\bibfield  {journal} {\bibinfo  {journal} {Phys. Rev. B}\
  }\textbf {\bibinfo {volume} {92}},\ \bibinfo {pages} {161110} (\bibinfo
  {year} {2015}{\natexlab{a}})}\BibitemShut {NoStop}%
\bibitem [{\citenamefont {Xu}\ \emph {et~al.}(2015{\natexlab{a}})\citenamefont
  {Xu}, \citenamefont {Belopolski}, \citenamefont {Alidoust}, \citenamefont
  {Neupane}, \citenamefont {Bian}, \citenamefont {Zhang}, \citenamefont
  {Sankar}, \citenamefont {Chang}, \citenamefont {Yuan}, \citenamefont {Lee},
  \citenamefont {Huang}, \citenamefont {Zheng}, \citenamefont {Ma},
  \citenamefont {Sanchez}, \citenamefont {Wang}, \citenamefont {Bansil},
  \citenamefont {Chou}, \citenamefont {Shibayev}, \citenamefont {Lin},
  \citenamefont {Jia},\ and\ \citenamefont {Hasan}}]{xu15}%
  \BibitemOpen
  \bibfield  {author} {\bibinfo {author} {\bibfnamefont {S.-Y.}\ \bibnamefont
  {Xu}}, \bibinfo {author} {\bibfnamefont {I.}~\bibnamefont {Belopolski}},
  \bibinfo {author} {\bibfnamefont {N.}~\bibnamefont {Alidoust}}, \bibinfo
  {author} {\bibfnamefont {M.}~\bibnamefont {Neupane}}, \bibinfo {author}
  {\bibfnamefont {G.}~\bibnamefont {Bian}}, \bibinfo {author} {\bibfnamefont
  {C.}~\bibnamefont {Zhang}}, \bibinfo {author} {\bibfnamefont
  {R.}~\bibnamefont {Sankar}}, \bibinfo {author} {\bibfnamefont
  {G.}~\bibnamefont {Chang}}, \bibinfo {author} {\bibfnamefont
  {Z.}~\bibnamefont {Yuan}}, \bibinfo {author} {\bibfnamefont {C.-C.}\
  \bibnamefont {Lee}}, \bibinfo {author} {\bibfnamefont {S.-M.}\ \bibnamefont
  {Huang}}, \bibinfo {author} {\bibfnamefont {H.}~\bibnamefont {Zheng}},
  \bibinfo {author} {\bibfnamefont {J.}~\bibnamefont {Ma}}, \bibinfo {author}
  {\bibfnamefont {D.~S.}\ \bibnamefont {Sanchez}}, \bibinfo {author}
  {\bibfnamefont {B.}~\bibnamefont {Wang}}, \bibinfo {author} {\bibfnamefont
  {A.}~\bibnamefont {Bansil}}, \bibinfo {author} {\bibfnamefont
  {F.}~\bibnamefont {Chou}}, \bibinfo {author} {\bibfnamefont {P.~P.}\
  \bibnamefont {Shibayev}}, \bibinfo {author} {\bibfnamefont {H.}~\bibnamefont
  {Lin}}, \bibinfo {author} {\bibfnamefont {S.}~\bibnamefont {Jia}}, \ and\
  \bibinfo {author} {\bibfnamefont {M.~Z.}\ \bibnamefont {Hasan}},\ }\href@noop
  {} {\bibfield  {journal} {\bibinfo  {journal} {Science}\ }\textbf {\bibinfo
  {volume} {349}},\ \bibinfo {pages} {613} (\bibinfo {year}
  {2015}{\natexlab{a}})}\BibitemShut {NoStop}%
\bibitem [{\citenamefont {Lv}\ \emph {et~al.}(2015)\citenamefont {Lv},
  \citenamefont {Weng}, \citenamefont {Fu}, \citenamefont {Wang}, \citenamefont
  {Miao}, \citenamefont {Ma}, \citenamefont {Richard}, \citenamefont {Huang},
  \citenamefont {Zhao}, \citenamefont {Chen}, \citenamefont {Fang},
  \citenamefont {Dai}, \citenamefont {Qian},\ and\ \citenamefont
  {Ding}}]{lv15}%
  \BibitemOpen
  \bibfield  {author} {\bibinfo {author} {\bibfnamefont {B.~Q.}\ \bibnamefont
  {Lv}}, \bibinfo {author} {\bibfnamefont {H.~M.}\ \bibnamefont {Weng}},
  \bibinfo {author} {\bibfnamefont {B.~B.}\ \bibnamefont {Fu}}, \bibinfo
  {author} {\bibfnamefont {X.~P.}\ \bibnamefont {Wang}}, \bibinfo {author}
  {\bibfnamefont {H.}~\bibnamefont {Miao}}, \bibinfo {author} {\bibfnamefont
  {J.}~\bibnamefont {Ma}}, \bibinfo {author} {\bibfnamefont {P.}~\bibnamefont
  {Richard}}, \bibinfo {author} {\bibfnamefont {X.~C.}\ \bibnamefont {Huang}},
  \bibinfo {author} {\bibfnamefont {L.~X.}\ \bibnamefont {Zhao}}, \bibinfo
  {author} {\bibfnamefont {G.~F.}\ \bibnamefont {Chen}}, \bibinfo {author}
  {\bibfnamefont {Z.}~\bibnamefont {Fang}}, \bibinfo {author} {\bibfnamefont
  {X.}~\bibnamefont {Dai}}, \bibinfo {author} {\bibfnamefont {T.}~\bibnamefont
  {Qian}}, \ and\ \bibinfo {author} {\bibfnamefont {H.}~\bibnamefont {Ding}},\
  }\href@noop {} {\bibfield  {journal} {\bibinfo  {journal} {Phys. Rev. X}\
  }\textbf {\bibinfo {volume} {5}},\ \bibinfo {pages} {031013} (\bibinfo {year}
  {2015})}\BibitemShut {NoStop}%
\bibitem [{\citenamefont {Xu}\ \emph {et~al.}(2015{\natexlab{b}})\citenamefont
  {Xu}, \citenamefont {Alidoust}, \citenamefont {Belopolski}, \citenamefont
  {Yuan}, \citenamefont {Bian}, \citenamefont {Chang}, \citenamefont {Zheng},
  \citenamefont {Strocov}, \citenamefont {Sanchez}, \citenamefont {Chang},
  \citenamefont {Zhang}, \citenamefont {Mou}, \citenamefont {Wu}, \citenamefont
  {Huang}, \citenamefont {Lee}, \citenamefont {Huang}, \citenamefont {Wang},
  \citenamefont {Bansil}, \citenamefont {Jeng}, \citenamefont {Neupert},
  \citenamefont {Kaminski}, \citenamefont {Lin}, \citenamefont {Jia},\ and\
  \citenamefont {Zahid~Hasan}}]{xu15b}%
  \BibitemOpen
  \bibfield  {author} {\bibinfo {author} {\bibfnamefont {S.-Y.}\ \bibnamefont
  {Xu}}, \bibinfo {author} {\bibfnamefont {N.}~\bibnamefont {Alidoust}},
  \bibinfo {author} {\bibfnamefont {I.}~\bibnamefont {Belopolski}}, \bibinfo
  {author} {\bibfnamefont {Z.}~\bibnamefont {Yuan}}, \bibinfo {author}
  {\bibfnamefont {G.}~\bibnamefont {Bian}}, \bibinfo {author} {\bibfnamefont
  {T.-R.}\ \bibnamefont {Chang}}, \bibinfo {author} {\bibfnamefont
  {H.}~\bibnamefont {Zheng}}, \bibinfo {author} {\bibfnamefont {V.~N.}\
  \bibnamefont {Strocov}}, \bibinfo {author} {\bibfnamefont {D.~S.}\
  \bibnamefont {Sanchez}}, \bibinfo {author} {\bibfnamefont {G.}~\bibnamefont
  {Chang}}, \bibinfo {author} {\bibfnamefont {C.}~\bibnamefont {Zhang}},
  \bibinfo {author} {\bibfnamefont {D.}~\bibnamefont {Mou}}, \bibinfo {author}
  {\bibfnamefont {Y.}~\bibnamefont {Wu}}, \bibinfo {author} {\bibfnamefont
  {L.}~\bibnamefont {Huang}}, \bibinfo {author} {\bibfnamefont {C.-C.}\
  \bibnamefont {Lee}}, \bibinfo {author} {\bibfnamefont {S.-M.}\ \bibnamefont
  {Huang}}, \bibinfo {author} {\bibfnamefont {B.}~\bibnamefont {Wang}},
  \bibinfo {author} {\bibfnamefont {A.}~\bibnamefont {Bansil}}, \bibinfo
  {author} {\bibfnamefont {H.-T.}\ \bibnamefont {Jeng}}, \bibinfo {author}
  {\bibfnamefont {T.}~\bibnamefont {Neupert}}, \bibinfo {author} {\bibfnamefont
  {A.}~\bibnamefont {Kaminski}}, \bibinfo {author} {\bibfnamefont
  {H.}~\bibnamefont {Lin}}, \bibinfo {author} {\bibfnamefont {S.}~\bibnamefont
  {Jia}}, \ and\ \bibinfo {author} {\bibfnamefont {M.}~\bibnamefont
  {Zahid~Hasan}},\ }\href@noop {} {\bibfield  {journal} {\bibinfo  {journal}
  {Nat Phys}\ }\textbf {\bibinfo {volume} {11}},\ \bibinfo {pages} {748}
  (\bibinfo {year} {2015}{\natexlab{b}})}\BibitemShut {NoStop}%
\bibitem [{\citenamefont {Borisenko}\ \emph {et~al.}(2015)\citenamefont
  {Borisenko}, \citenamefont {Evtushinsky}, \citenamefont {Gibson},
  \citenamefont {Yaresko}, \citenamefont {Kim}, \citenamefont {Ali},
  \citenamefont {Buechner}, \citenamefont {Hoesch},\ and\ \citenamefont
  {Cava}}]{borisenko15}%
  \BibitemOpen
  \bibfield  {author} {\bibinfo {author} {\bibfnamefont {S.}~\bibnamefont
  {Borisenko}}, \bibinfo {author} {\bibfnamefont {D.}~\bibnamefont
  {Evtushinsky}}, \bibinfo {author} {\bibfnamefont {Q.}~\bibnamefont {Gibson}},
  \bibinfo {author} {\bibfnamefont {A.}~\bibnamefont {Yaresko}}, \bibinfo
  {author} {\bibfnamefont {T.}~\bibnamefont {Kim}}, \bibinfo {author}
  {\bibfnamefont {M.~N.}\ \bibnamefont {Ali}}, \bibinfo {author} {\bibfnamefont
  {B.}~\bibnamefont {Buechner}}, \bibinfo {author} {\bibfnamefont
  {M.}~\bibnamefont {Hoesch}}, \ and\ \bibinfo {author} {\bibfnamefont {R.~J.}\
  \bibnamefont {Cava}},\ }\href@noop {} {\bibfield  {journal} {\bibinfo
  {journal} {arXiv:1507.04847}\ } (\bibinfo {year} {2015})}\BibitemShut
  {NoStop}%
\bibitem [{\citenamefont {Sushkov}\ \emph {et~al.}(2015)\citenamefont
  {Sushkov}, \citenamefont {Hofmann}, \citenamefont {Jenkins}, \citenamefont
  {Ishikawa}, \citenamefont {Nakatsuji}, \citenamefont {Das~Sarma},\ and\
  \citenamefont {Drew}}]{sushkov15}%
  \BibitemOpen
  \bibfield  {author} {\bibinfo {author} {\bibfnamefont {A.~B.}\ \bibnamefont
  {Sushkov}}, \bibinfo {author} {\bibfnamefont {J.~B.}\ \bibnamefont
  {Hofmann}}, \bibinfo {author} {\bibfnamefont {G.~S.}\ \bibnamefont
  {Jenkins}}, \bibinfo {author} {\bibfnamefont {J.}~\bibnamefont {Ishikawa}},
  \bibinfo {author} {\bibfnamefont {S.}~\bibnamefont {Nakatsuji}}, \bibinfo
  {author} {\bibfnamefont {S.}~\bibnamefont {Das~Sarma}}, \ and\ \bibinfo
  {author} {\bibfnamefont {H.~D.}\ \bibnamefont {Drew}},\ }\href@noop {}
  {\bibfield  {journal} {\bibinfo  {journal} {Phys. Rev. B}\ }\textbf {\bibinfo
  {volume} {92}},\ \bibinfo {pages} {241108} (\bibinfo {year}
  {2015})}\BibitemShut {NoStop}%
\bibitem [{\citenamefont {Borisenko}\ \emph {et~al.}(2014)\citenamefont
  {Borisenko}, \citenamefont {Gibson}, \citenamefont {Evtushinsky},
  \citenamefont {Zabolotnyy}, \citenamefont {B\"uchner},\ and\ \citenamefont
  {Cava}}]{borisenko14}%
  \BibitemOpen
  \bibfield  {author} {\bibinfo {author} {\bibfnamefont {S.}~\bibnamefont
  {Borisenko}}, \bibinfo {author} {\bibfnamefont {Q.}~\bibnamefont {Gibson}},
  \bibinfo {author} {\bibfnamefont {D.}~\bibnamefont {Evtushinsky}}, \bibinfo
  {author} {\bibfnamefont {V.}~\bibnamefont {Zabolotnyy}}, \bibinfo {author}
  {\bibfnamefont {B.}~\bibnamefont {B\"uchner}}, \ and\ \bibinfo {author}
  {\bibfnamefont {R.~J.}\ \bibnamefont {Cava}},\ }\href@noop {} {\bibfield
  {journal} {\bibinfo  {journal} {Phys. Rev. Lett.}\ }\textbf {\bibinfo
  {volume} {113}},\ \bibinfo {pages} {027603} (\bibinfo {year}
  {2014})}\BibitemShut {NoStop}%
\bibitem [{\citenamefont {Liang}\ \emph {et~al.}(2015)\citenamefont {Liang},
  \citenamefont {Gibson}, \citenamefont {Ali}, \citenamefont {Liu},
  \citenamefont {Cava},\ and\ \citenamefont {Ong}}]{liang15}%
  \BibitemOpen
  \bibfield  {author} {\bibinfo {author} {\bibfnamefont {T.}~\bibnamefont
  {Liang}}, \bibinfo {author} {\bibfnamefont {Q.}~\bibnamefont {Gibson}},
  \bibinfo {author} {\bibfnamefont {M.~N.}\ \bibnamefont {Ali}}, \bibinfo
  {author} {\bibfnamefont {M.}~\bibnamefont {Liu}}, \bibinfo {author}
  {\bibfnamefont {R.~J.}\ \bibnamefont {Cava}}, \ and\ \bibinfo {author}
  {\bibfnamefont {N.~P.}\ \bibnamefont {Ong}},\ }\href@noop {} {\bibfield
  {journal} {\bibinfo  {journal} {Nat Mater}\ }\textbf {\bibinfo {volume}
  {14}},\ \bibinfo {pages} {280} (\bibinfo {year} {2015})}\BibitemShut
  {NoStop}%
\bibitem [{\citenamefont {Neupane}\ \emph {et~al.}(2014)\citenamefont
  {Neupane}, \citenamefont {Xu}, \citenamefont {Sankar}, \citenamefont
  {Alidoust}, \citenamefont {Bian}, \citenamefont {Liu}, \citenamefont
  {Belopolski}, \citenamefont {Chang}, \citenamefont {Jeng}, \citenamefont
  {Lin}, \citenamefont {Bansil}, \citenamefont {Chou},\ and\ \citenamefont
  {Hasan}}]{neupane14}%
  \BibitemOpen
  \bibfield  {author} {\bibinfo {author} {\bibfnamefont {M.}~\bibnamefont
  {Neupane}}, \bibinfo {author} {\bibfnamefont {S.-Y.}\ \bibnamefont {Xu}},
  \bibinfo {author} {\bibfnamefont {R.}~\bibnamefont {Sankar}}, \bibinfo
  {author} {\bibfnamefont {N.}~\bibnamefont {Alidoust}}, \bibinfo {author}
  {\bibfnamefont {G.}~\bibnamefont {Bian}}, \bibinfo {author} {\bibfnamefont
  {C.}~\bibnamefont {Liu}}, \bibinfo {author} {\bibfnamefont {I.}~\bibnamefont
  {Belopolski}}, \bibinfo {author} {\bibfnamefont {T.-R.}\ \bibnamefont
  {Chang}}, \bibinfo {author} {\bibfnamefont {H.-T.}\ \bibnamefont {Jeng}},
  \bibinfo {author} {\bibfnamefont {H.}~\bibnamefont {Lin}}, \bibinfo {author}
  {\bibfnamefont {A.}~\bibnamefont {Bansil}}, \bibinfo {author} {\bibfnamefont
  {F.}~\bibnamefont {Chou}}, \ and\ \bibinfo {author} {\bibfnamefont {M.~Z.}\
  \bibnamefont {Hasan}},\ }\href@noop {} {\bibfield  {journal} {\bibinfo
  {journal} {Nat Commun}\ }\textbf {\bibinfo {volume} {5}},\ \bibinfo {pages}
  {3786} (\bibinfo {year} {2014})}\BibitemShut {NoStop}%
\bibitem [{\citenamefont {Li}\ \emph {et~al.}(2014)\citenamefont {Li},
  \citenamefont {Kharzeev}, \citenamefont {Zhang}, \citenamefont {Huang},
  \citenamefont {Pletikosic}, \citenamefont {Fedorov}, \citenamefont {Zhong},
  \citenamefont {Schneeloch}, \citenamefont {Gu},\ and\ \citenamefont
  {Valla}}]{li14}%
  \BibitemOpen
  \bibfield  {author} {\bibinfo {author} {\bibfnamefont {Q.}~\bibnamefont
  {Li}}, \bibinfo {author} {\bibfnamefont {D.~E.}\ \bibnamefont {Kharzeev}},
  \bibinfo {author} {\bibfnamefont {C.}~\bibnamefont {Zhang}}, \bibinfo
  {author} {\bibfnamefont {Y.}~\bibnamefont {Huang}}, \bibinfo {author}
  {\bibfnamefont {I.}~\bibnamefont {Pletikosic}}, \bibinfo {author}
  {\bibfnamefont {A.~V.}\ \bibnamefont {Fedorov}}, \bibinfo {author}
  {\bibfnamefont {R.~D.}\ \bibnamefont {Zhong}}, \bibinfo {author}
  {\bibfnamefont {J.~A.}\ \bibnamefont {Schneeloch}}, \bibinfo {author}
  {\bibfnamefont {G.~D.}\ \bibnamefont {Gu}}, \ and\ \bibinfo {author}
  {\bibfnamefont {T.}~\bibnamefont {Valla}},\ }\href@noop {} {\bibfield
  {journal} {\bibinfo  {journal} {arXiv:1412.6543}\ } (\bibinfo {year}
  {2014})}\BibitemShut {NoStop}%
\bibitem [{\citenamefont {Liu}\ \emph {et~al.}(2014)\citenamefont {Liu},
  \citenamefont {Zhou}, \citenamefont {Zhang}, \citenamefont {Wang},
  \citenamefont {Weng}, \citenamefont {Prabhakaran}, \citenamefont {Mo},
  \citenamefont {Shen}, \citenamefont {Fang}, \citenamefont {Dai},
  \citenamefont {Hussain},\ and\ \citenamefont {Chen}}]{liu14}%
  \BibitemOpen
  \bibfield  {author} {\bibinfo {author} {\bibfnamefont {Z.~K.}\ \bibnamefont
  {Liu}}, \bibinfo {author} {\bibfnamefont {B.}~\bibnamefont {Zhou}}, \bibinfo
  {author} {\bibfnamefont {Y.}~\bibnamefont {Zhang}}, \bibinfo {author}
  {\bibfnamefont {Z.~J.}\ \bibnamefont {Wang}}, \bibinfo {author}
  {\bibfnamefont {H.~M.}\ \bibnamefont {Weng}}, \bibinfo {author}
  {\bibfnamefont {D.}~\bibnamefont {Prabhakaran}}, \bibinfo {author}
  {\bibfnamefont {S.-K.}\ \bibnamefont {Mo}}, \bibinfo {author} {\bibfnamefont
  {Z.~X.}\ \bibnamefont {Shen}}, \bibinfo {author} {\bibfnamefont
  {Z.}~\bibnamefont {Fang}}, \bibinfo {author} {\bibfnamefont {X.}~\bibnamefont
  {Dai}}, \bibinfo {author} {\bibfnamefont {Z.}~\bibnamefont {Hussain}}, \ and\
  \bibinfo {author} {\bibfnamefont {Y.~L.}\ \bibnamefont {Chen}},\ }\href@noop
  {} {\bibfield  {journal} {\bibinfo  {journal} {Science}\ }\textbf {\bibinfo
  {volume} {343}},\ \bibinfo {pages} {864} (\bibinfo {year}
  {2014})}\BibitemShut {NoStop}%
\bibitem [{\citenamefont {Kim}\ \emph {et~al.}(2013)\citenamefont {Kim},
  \citenamefont {Kim}, \citenamefont {Wang}, \citenamefont {Sasaki},
  \citenamefont {Satoh}, \citenamefont {Ohnishi}, \citenamefont {Kitaura},
  \citenamefont {Yang},\ and\ \citenamefont {Li}}]{kim13}%
  \BibitemOpen
  \bibfield  {author} {\bibinfo {author} {\bibfnamefont {H.-J.}\ \bibnamefont
  {Kim}}, \bibinfo {author} {\bibfnamefont {K.-S.}\ \bibnamefont {Kim}},
  \bibinfo {author} {\bibfnamefont {J.-F.}\ \bibnamefont {Wang}}, \bibinfo
  {author} {\bibfnamefont {M.}~\bibnamefont {Sasaki}}, \bibinfo {author}
  {\bibfnamefont {N.}~\bibnamefont {Satoh}}, \bibinfo {author} {\bibfnamefont
  {A.}~\bibnamefont {Ohnishi}}, \bibinfo {author} {\bibfnamefont
  {M.}~\bibnamefont {Kitaura}}, \bibinfo {author} {\bibfnamefont
  {M.}~\bibnamefont {Yang}}, \ and\ \bibinfo {author} {\bibfnamefont
  {L.}~\bibnamefont {Li}},\ }\href@noop {} {\bibfield  {journal} {\bibinfo
  {journal} {Phys. Rev. Lett.}\ }\textbf {\bibinfo {volume} {111}},\ \bibinfo
  {pages} {246603} (\bibinfo {year} {2013})}\BibitemShut {NoStop}%
\bibitem [{\citenamefont {Xiong}\ \emph {et~al.}(2015)\citenamefont {Xiong},
  \citenamefont {Kushwaha}, \citenamefont {Liang}, \citenamefont {Krizan},
  \citenamefont {Wang}, \citenamefont {Cava},\ and\ \citenamefont
  {Ong}}]{xiong15}%
  \BibitemOpen
  \bibfield  {author} {\bibinfo {author} {\bibfnamefont {J.}~\bibnamefont
  {Xiong}}, \bibinfo {author} {\bibfnamefont {S.~K.}\ \bibnamefont {Kushwaha}},
  \bibinfo {author} {\bibfnamefont {T.}~\bibnamefont {Liang}}, \bibinfo
  {author} {\bibfnamefont {J.~W.}\ \bibnamefont {Krizan}}, \bibinfo {author}
  {\bibfnamefont {W.}~\bibnamefont {Wang}}, \bibinfo {author} {\bibfnamefont
  {R.~J.}\ \bibnamefont {Cava}}, \ and\ \bibinfo {author} {\bibfnamefont
  {N.~P.}\ \bibnamefont {Ong}},\ }\href@noop {} {\bibfield  {journal} {\bibinfo
   {journal} {arXiv:1503.08179}\ } (\bibinfo {year} {2015})}\BibitemShut
  {NoStop}%
\bibitem [{\citenamefont {Goswami}\ \emph
  {et~al.}(2015{\natexlab{b}})\citenamefont {Goswami}, \citenamefont {Pixley},\
  and\ \citenamefont {Das~Sarma}}]{goswami15}%
  \BibitemOpen
  \bibfield  {author} {\bibinfo {author} {\bibfnamefont {P.}~\bibnamefont
  {Goswami}}, \bibinfo {author} {\bibfnamefont {J.~H.}\ \bibnamefont {Pixley}},
  \ and\ \bibinfo {author} {\bibfnamefont {S.}~\bibnamefont {Das~Sarma}},\
  }\href@noop {} {\bibfield  {journal} {\bibinfo  {journal} {Phys. Rev. B}\
  }\textbf {\bibinfo {volume} {92}},\ \bibinfo {pages} {075205} (\bibinfo
  {year} {2015}{\natexlab{b}})}\BibitemShut {NoStop}%
\bibitem [{\citenamefont {Son}\ and\ \citenamefont {Spivak}(2013)}]{son13}%
  \BibitemOpen
  \bibfield  {author} {\bibinfo {author} {\bibfnamefont {D.~T.}\ \bibnamefont
  {Son}}\ and\ \bibinfo {author} {\bibfnamefont {B.~Z.}\ \bibnamefont
  {Spivak}},\ }\href@noop {} {\bibfield  {journal} {\bibinfo  {journal} {Phys.
  Rev. B}\ }\textbf {\bibinfo {volume} {88}},\ \bibinfo {pages} {104412}
  (\bibinfo {year} {2013})}\BibitemShut {NoStop}%
\bibitem [{\citenamefont {Spivak}\ and\ \citenamefont
  {Andreev}(2015)}]{spivak15}%
  \BibitemOpen
  \bibfield  {author} {\bibinfo {author} {\bibfnamefont {B.~Z.}\ \bibnamefont
  {Spivak}}\ and\ \bibinfo {author} {\bibfnamefont {A.~V.}\ \bibnamefont
  {Andreev}},\ }\href@noop {} {\bibfield  {journal} {\bibinfo  {journal}
  {arXiv:1510.01817}\ } (\bibinfo {year} {2015})}\BibitemShut {NoStop}%
\bibitem [{\citenamefont {Das~Sarma}\ \emph {et~al.}(2015)\citenamefont
  {Das~Sarma}, \citenamefont {Hwang},\ and\ \citenamefont {Min}}]{dassarma15}%
  \BibitemOpen
  \bibfield  {author} {\bibinfo {author} {\bibfnamefont {S.}~\bibnamefont
  {Das~Sarma}}, \bibinfo {author} {\bibfnamefont {E.~H.}\ \bibnamefont
  {Hwang}}, \ and\ \bibinfo {author} {\bibfnamefont {H.}~\bibnamefont {Min}},\
  }\href@noop {} {\bibfield  {journal} {\bibinfo  {journal} {Phys. Rev. B}\
  }\textbf {\bibinfo {volume} {91}},\ \bibinfo {pages} {035201} (\bibinfo
  {year} {2015})}\BibitemShut {NoStop}%
\bibitem [{\citenamefont {Li}\ and\ \citenamefont {Andreev}(2015)}]{li15}%
  \BibitemOpen
  \bibfield  {author} {\bibinfo {author} {\bibfnamefont {S.}~\bibnamefont
  {Li}}\ and\ \bibinfo {author} {\bibfnamefont {A.~V.}\ \bibnamefont
  {Andreev}},\ }\href@noop {} {\bibfield  {journal} {\bibinfo  {journal} {Phys.
  Rev. B}\ }\textbf {\bibinfo {volume} {92}},\ \bibinfo {pages} {201107}
  (\bibinfo {year} {2015})}\BibitemShut {NoStop}%
\bibitem [{\citenamefont {Lv}\ and\ \citenamefont {Zhang}(2013)}]{lv13}%
  \BibitemOpen
  \bibfield  {author} {\bibinfo {author} {\bibfnamefont {M.}~\bibnamefont
  {Lv}}\ and\ \bibinfo {author} {\bibfnamefont {S.-C.}\ \bibnamefont {Zhang}},\
  }\href@noop {} {\bibfield  {journal} {\bibinfo  {journal} {International
  Journal of Modern Physics B}\ }\textbf {\bibinfo {volume} {27}},\ \bibinfo
  {pages} {1350177} (\bibinfo {year} {2013})}\BibitemShut {NoStop}%
\bibitem [{\citenamefont {Hofmann}\ and\ \citenamefont
  {Das~Sarma}(2015)}]{hofmann15}%
  \BibitemOpen
  \bibfield  {author} {\bibinfo {author} {\bibfnamefont {J.}~\bibnamefont
  {Hofmann}}\ and\ \bibinfo {author} {\bibfnamefont {S.}~\bibnamefont
  {Das~Sarma}},\ }\href@noop {} {\bibfield  {journal} {\bibinfo  {journal}
  {Phys. Rev. B}\ }\textbf {\bibinfo {volume} {91}},\ \bibinfo {pages} {241108}
  (\bibinfo {year} {2015})}\BibitemShut {NoStop}%
\bibitem [{\citenamefont {Zhou}\ \emph {et~al.}(2015)\citenamefont {Zhou},
  \citenamefont {Chang},\ and\ \citenamefont {Xiao}}]{zhou15}%
  \BibitemOpen
  \bibfield  {author} {\bibinfo {author} {\bibfnamefont {J.}~\bibnamefont
  {Zhou}}, \bibinfo {author} {\bibfnamefont {H.-R.}\ \bibnamefont {Chang}}, \
  and\ \bibinfo {author} {\bibfnamefont {D.}~\bibnamefont {Xiao}},\ }\href@noop
  {} {\bibfield  {journal} {\bibinfo  {journal} {Phys. Rev. B}\ }\textbf
  {\bibinfo {volume} {91}},\ \bibinfo {pages} {035114} (\bibinfo {year}
  {2015})}\BibitemShut {NoStop}%
\bibitem [{\citenamefont {Pitarke}\ \emph {et~al.}(2007)\citenamefont
  {Pitarke}, \citenamefont {Silkin}, \citenamefont {Chulkov},\ and\
  \citenamefont {Echenique}}]{pitarke07}%
  \BibitemOpen
  \bibfield  {author} {\bibinfo {author} {\bibfnamefont {J.~M.}\ \bibnamefont
  {Pitarke}}, \bibinfo {author} {\bibfnamefont {V.~M.}\ \bibnamefont {Silkin}},
  \bibinfo {author} {\bibfnamefont {E.~V.}\ \bibnamefont {Chulkov}}, \ and\
  \bibinfo {author} {\bibfnamefont {P.~M.}\ \bibnamefont {Echenique}},\
  }\href@noop {} {\bibfield  {journal} {\bibinfo  {journal} {Reports on
  Progress in Physics}\ }\textbf {\bibinfo {volume} {70}},\ \bibinfo {pages}
  {1} (\bibinfo {year} {2007})}\BibitemShut {NoStop}%
\bibitem [{\citenamefont {Das~Sarma}\ and\ \citenamefont
  {Hwang}(2009)}]{dassarma09}%
  \BibitemOpen
  \bibfield  {author} {\bibinfo {author} {\bibfnamefont {S.}~\bibnamefont
  {Das~Sarma}}\ and\ \bibinfo {author} {\bibfnamefont {E.~H.}\ \bibnamefont
  {Hwang}},\ }\href@noop {} {\bibfield  {journal} {\bibinfo  {journal} {Phys.
  Rev. Lett.}\ }\textbf {\bibinfo {volume} {102}},\ \bibinfo {pages} {206412}
  (\bibinfo {year} {2009})}\BibitemShut {NoStop}%
\bibitem [{\citenamefont {Wallis}\ \emph {et~al.}(1974)\citenamefont {Wallis},
  \citenamefont {Brion}, \citenamefont {Burstein},\ and\ \citenamefont
  {Hartstein}}]{wallis74}%
  \BibitemOpen
  \bibfield  {author} {\bibinfo {author} {\bibfnamefont {R.~F.}\ \bibnamefont
  {Wallis}}, \bibinfo {author} {\bibfnamefont {J.~J.}\ \bibnamefont {Brion}},
  \bibinfo {author} {\bibfnamefont {E.}~\bibnamefont {Burstein}}, \ and\
  \bibinfo {author} {\bibfnamefont {A.}~\bibnamefont {Hartstein}},\ }\href@noop
  {} {\bibfield  {journal} {\bibinfo  {journal} {Phys. Rev. B}\ }\textbf
  {\bibinfo {volume} {9}},\ \bibinfo {pages} {3424} (\bibinfo {year}
  {1974})}\BibitemShut {NoStop}%
\bibitem [{\citenamefont {Camley}(1987)}]{camley87}%
  \BibitemOpen
  \bibfield  {author} {\bibinfo {author} {\bibfnamefont {R.}~\bibnamefont
  {Camley}},\ }\href@noop {} {\bibfield  {journal} {\bibinfo  {journal}
  {Surface Science Reports}\ }\textbf {\bibinfo {volume} {7}},\ \bibinfo
  {pages} {103 } (\bibinfo {year} {1987})}\BibitemShut {NoStop}%
\bibitem [{\citenamefont {Hartstein}\ and\ \citenamefont
  {Burstein}(1974)}]{hartstein74}%
  \BibitemOpen
  \bibfield  {author} {\bibinfo {author} {\bibfnamefont {A.}~\bibnamefont
  {Hartstein}}\ and\ \bibinfo {author} {\bibfnamefont {E.}~\bibnamefont
  {Burstein}},\ }\href@noop {} {\bibfield  {journal} {\bibinfo  {journal}
  {Solid State Communications}\ }\textbf {\bibinfo {volume} {14}},\ \bibinfo
  {pages} {1223 } (\bibinfo {year} {1974})}\BibitemShut {NoStop}%
\bibitem [{\citenamefont {Chiu}\ and\ \citenamefont {Quinn}(1972)}]{chiu72}%
  \BibitemOpen
  \bibfield  {author} {\bibinfo {author} {\bibfnamefont {K.}~\bibnamefont
  {Chiu}}\ and\ \bibinfo {author} {\bibfnamefont {J.}~\bibnamefont {Quinn}},\
  }\href@noop {} {\bibfield  {journal} {\bibinfo  {journal} {Il Nuovo Cimento
  B}\ }\textbf {\bibinfo {volume} {10}},\ \bibinfo {pages} {1} (\bibinfo {year}
  {1972})}\BibitemShut {NoStop}%
\bibitem [{\citenamefont {Kushwaha}(2001)}]{kushwaha01}%
  \BibitemOpen
  \bibfield  {author} {\bibinfo {author} {\bibfnamefont {M.~S.}\ \bibnamefont
  {Kushwaha}},\ }\href@noop {} {\bibfield  {journal} {\bibinfo  {journal}
  {Surface Science Reports}\ }\textbf {\bibinfo {volume} {41}},\ \bibinfo
  {pages} {1 } (\bibinfo {year} {2001})}\BibitemShut {NoStop}%
\bibitem [{\citenamefont {Karch}(2011)}]{karch11}%
  \BibitemOpen
  \bibfield  {author} {\bibinfo {author} {\bibfnamefont {A.}~\bibnamefont
  {Karch}},\ }\href@noop {} {\bibfield  {journal} {\bibinfo  {journal} {Phys.
  Rev. B}\ }\textbf {\bibinfo {volume} {83}},\ \bibinfo {pages} {245432}
  (\bibinfo {year} {2011})}\BibitemShut {NoStop}%
\bibitem [{\citenamefont {Xiao}\ \emph {et~al.}(2010)\citenamefont {Xiao},
  \citenamefont {Chang},\ and\ \citenamefont {Niu}}]{xiao10}%
  \BibitemOpen
  \bibfield  {author} {\bibinfo {author} {\bibfnamefont {D.}~\bibnamefont
  {Xiao}}, \bibinfo {author} {\bibfnamefont {M.-C.}\ \bibnamefont {Chang}}, \
  and\ \bibinfo {author} {\bibfnamefont {Q.}~\bibnamefont {Niu}},\ }\href@noop
  {} {\bibfield  {journal} {\bibinfo  {journal} {Rev. Mod. Phys.}\ }\textbf
  {\bibinfo {volume} {82}},\ \bibinfo {pages} {1959} (\bibinfo {year}
  {2010})}\BibitemShut {NoStop}%
\bibitem [{\citenamefont {Lundgren}\ \emph {et~al.}(2014)\citenamefont
  {Lundgren}, \citenamefont {Laurell},\ and\ \citenamefont
  {Fiete}}]{lundgren14}%
  \BibitemOpen
  \bibfield  {author} {\bibinfo {author} {\bibfnamefont {R.}~\bibnamefont
  {Lundgren}}, \bibinfo {author} {\bibfnamefont {P.}~\bibnamefont {Laurell}}, \
  and\ \bibinfo {author} {\bibfnamefont {G.~A.}\ \bibnamefont {Fiete}},\
  }\href@noop {} {\bibfield  {journal} {\bibinfo  {journal} {Phys. Rev. B}\
  }\textbf {\bibinfo {volume} {90}},\ \bibinfo {pages} {165115} (\bibinfo
  {year} {2014})}\BibitemShut {NoStop}%
\bibitem [{\citenamefont {Fetter}(1985)}]{fetter85}%
  \BibitemOpen
  \bibfield  {author} {\bibinfo {author} {\bibfnamefont {A.~L.}\ \bibnamefont
  {Fetter}},\ }\href@noop {} {\bibfield  {journal} {\bibinfo  {journal} {Phys.
  Rev. B}\ }\textbf {\bibinfo {volume} {32}},\ \bibinfo {pages} {7676}
  (\bibinfo {year} {1985})}\BibitemShut {NoStop}%
\end{thebibliography}

\begin{thebibliography}{2}%
\makeatletter
\providecommand \@ifxundefined [1]{%
 \@ifx{#1\undefined}
}%
\providecommand \@ifnum [1]{%
 \ifnum #1\expandafter \@firstoftwo
 \else \expandafter \@secondoftwo
 \fi
}%
\providecommand \@ifx [1]{%
 \ifx #1\expandafter \@firstoftwo
 \else \expandafter \@secondoftwo
 \fi
}%
\providecommand \natexlab [1]{#1}%
\providecommand \enquote  [1]{``#1''}%
\providecommand \bibnamefont  [1]{#1}%
\providecommand \bibfnamefont [1]{#1}%
\providecommand \citenamefont [1]{#1}%
\providecommand \href@noop [0]{\@secondoftwo}%
\providecommand \href [0]{\begingroup \@sanitize@url \@href}%
\providecommand \@href[1]{\@@startlink{#1}\@@href}%
\providecommand \@@href[1]{\endgroup#1\@@endlink}%
\providecommand \@sanitize@url [0]{\catcode `\\12\catcode `\$12\catcode
  `\&12\catcode `\#12\catcode `\^12\catcode `\_12\catcode `\%12\relax}%
\providecommand \@@startlink[1]{}%
\providecommand \@@endlink[0]{}%
\providecommand \url  [0]{\begingroup\@sanitize@url \@url }%
\providecommand \@url [1]{\endgroup\@href {#1}{\urlprefix }}%
\providecommand \urlprefix  [0]{URL }%
\providecommand \Eprint [0]{\href }%
\providecommand \doibase [0]{http://dx.doi.org/}%
\providecommand \selectlanguage [0]{\@gobble}%
\providecommand \bibinfo  [0]{\@secondoftwo}%
\providecommand \bibfield  [0]{\@secondoftwo}%
\providecommand \translation [1]{[#1]}%
\providecommand \BibitemOpen [0]{}%
\providecommand \bibitemStop [0]{}%
\providecommand \bibitemNoStop [0]{.\EOS\space}%
\providecommand \EOS [0]{\spacefactor3000\relax}%
\providecommand \BibitemShut  [1]{\csname bibitem#1\endcsname}%
\let\auto@bib@innerbib\@empty
%</preamble>
\bibitem [{\citenamefont {Chiu}\ and\ \citenamefont
  {Quinn}(1972)}]{suppchiu72}%
  \BibitemOpen
  \bibfield  {author} {\bibinfo {author} {\bibfnamefont {K.}~\bibnamefont
  {Chiu}}\ and\ \bibinfo {author} {\bibfnamefont {J.}~\bibnamefont {Quinn}},\
  }\href@noop {} {\bibfield  {journal} {\bibinfo  {journal} {Il Nuovo Cimento
  B}\ }\textbf {\bibinfo {volume} {10}},\ \bibinfo {pages} {1} (\bibinfo {year}
  {1972})}\BibitemShut {NoStop}%
\bibitem [{\citenamefont {Wallis}\ \emph {et~al.}(1974)\citenamefont {Wallis},
  \citenamefont {Brion}, \citenamefont {Burstein},\ and\ \citenamefont
  {Hartstein}}]{suppwallis74}%
  \BibitemOpen
  \bibfield  {author} {\bibinfo {author} {\bibfnamefont {R.~F.}\ \bibnamefont
  {Wallis}}, \bibinfo {author} {\bibfnamefont {J.~J.}\ \bibnamefont {Brion}},
  \bibinfo {author} {\bibfnamefont {E.}~\bibnamefont {Burstein}}, \ and\
  \bibinfo {author} {\bibfnamefont {A.}~\bibnamefont {Hartstein}},\ }\href@noop
  {} {\bibfield  {journal} {\bibinfo  {journal} {Phys. Rev. B}\ }\textbf
  {\bibinfo {volume} {9}},\ \bibinfo {pages} {3424} (\bibinfo {year}
  {1974})}\BibitemShut {NoStop}%
\end{thebibliography}
\end{document}